\documentclass[12pt,a4]{article}
\pdfoutput=1
\usepackage[utf8]{inputenc}
\usepackage[labelfont=bf,textfont=it]{caption}
\usepackage{a4wide}
\usepackage{xspace}
\usepackage{amsmath,amssymb,slashed}
\usepackage{booktabs}
\usepackage{graphicx}
\usepackage{url}
\usepackage[breaklinks=true]{hyperref}
\usepackage{subfigure}
\usepackage{hhline,multirow}
\usepackage{adjustbox,array}
\usepackage{cite}
\usepackage{placeins}
\allowdisplaybreaks
\usepackage{enumitem}

\usepackage{xcolor}
\definecolor{comment}{rgb}{0,0.3,0}
\definecolor{identifier}{rgb}{0.0,0,0.3}
\usepackage{listings}
\newcommand{\be}{\begin{eqnarray*}}
\newcommand{\ee}{\end{eqnarray*}}

\newcommand{\bee}{\begin{eqnarray}}
\newcommand{\eee}{\end{eqnarray}}
\newcommand{\beeq}{\begin{equation}}
\newcommand{\eeeq}{\end{equation}}
\newcommand{\gev}{\ensuremath{\text{G}\mspace{0.2mu}\text{e}\mspace{-1mu}\text{V}}\xspace}
\newcommand{\tev}{\ensuremath{\text{T}\mspace{0.05mu}\text{e}\mspace{-1mu}\text{V}}\xspace}

\newcommand{\mc}{\mathcal}

\newcommand{\Leff}{\ensuremath{\mathcal{L}_\mathrm{eff}}\xspace}
\newcommand{\Lsm}{\ensuremath{\mathcal{L}_\mathrm{SM}}\xspace}
\newcommand{\Afb}{\ensuremath{A_\mathrm{FB}}\xspace}
\newcommand{\Ac}{\ensuremath{A_\mathrm{C}}\xspace}

\newcommand{\dzero}{\ensuremath{\text{D${\slashed{0}}$}}\xspace}

\newenvironment{cedescription}{%
   
   \begin{description}[leftmargin=0.25cm, style=sameline]%
}{%
   \end{description}%
}

\title{\vspace{2cm} \LARGE{\textbf{Constraining top quark effective theory\\ in the LHC Run~II era \vspace{1cm} }}
}

\author{The \textsc{TopFitter} Collaboration \vspace{1cm}\\
 Andy Buckley,~
 Christoph Englert,~
 James Ferrando,~
 David J. Miller,~\\
 Liam Moore,~
 Michael Russell,~
 and Chris D. White
 \vspace{1cm} \\
 \textit{SUPA, School of Physics and Astronomy, University of Glasgow,} \\
 \textit{Glasgow, G12 8QQ, United Kingdom}
}

\date{}

\begin{document}

\maketitle

\vspace{-15cm}
\begin{flushright}
  GLAS-PPE/2015-08
\end{flushright}
\vspace{15cm}

\thispagestyle{empty}

\begin{abstract}
  We perform an up-to-date global fit of top quark effective theory to
  experimental data from the Tevatron, and from LHC Runs~I and~II. Experimental
  data includes total cross-sections up to 13~\tev, as well as differential
  distributions, for both single top and pair production. We also include the
  top quark width, charge asymmetries, and polarisation information from top
  decay products. We present bounds on the coefficients of dimension six
  operators, and examine the interplay between inclusive and differential
  measurements, and Tevatron / LHC data. All results are currently in good
  agreement with the Standard~Model.
\end{abstract}

\newpage

\section{Introduction}
One of the primary goals of the Large Hadron Collider (LHC) is to uncover the
precise mechanism responsible for electroweak symmetry breaking. Going beyond its ad hoc implementation in the Standard Model (SM), most
realisations of this mechanism predict that new, possibly non-resonant physics will appear at the
(multi-)\tev scale. Faced with the large number of such scenarios, and the
frequent degeneracy in their experimental signatures, it has become customary to
parametrize deviations of LHC measurements from their Standard Model predictions
in terms of \textit{model-independent} parameters, where possible. In Higgs
production, for instance, the deviations in early inclusive cross-section
measurements are described by `signal strength' ratios. Likewise, deviations in electroweak parameters are often expressed in the
language of anomalous couplings.

With the LHC Run~I at a close, the main message to be drawn is that, apart from
a few scattered anomalies, all measurements are in agreement with
Standard Model predictions. This suggests that the new degrees of freedom, if
they exist at all, are separated in mass~\cite{Appelquist:1974tg,Wilson:1983dy} from the
Standard Model fields\footnote{Current collider measurements, however, cannot
  rule out the existence of light degrees of freedom, see e.g.
  Ref.~\cite{King:2012tr}.}. If this is true, the new physics can be modelled by
an infinite series of higher-dimensional effective
operators~\cite{Buchmuller:1985jz,Hagiwara:1986vm,Burges:1983zg,Leung:1984ni}. From a
phenomenological perspective, these have the advantage over simple signal
strengths in that they can also accommodate differential measurements and
angular observables, since the operators lead to new vertex structures which
modify event kinematics. They are also preferable to anomalous couplings since
they preserve the Standard Model $SU(3)_C\times SU(2)_L \times U(1)_Y$ gauge
symmetry, so can more easily be linked to ultraviolet completions than arbitrary
form factors. These merits have not gone unnoticed, as effective field theory
(EFT) techniques have received much attention in interpreting available Higgs
results~\cite{Azatov:2012bz,Espinosa:2012im,Plehn:2012iz,Carmi:2012in,Peskin:2012we,Dumont:2013wma,Djouadi:2013qya,Lopez-Val:2013yba,Englert:2014uua,Ellis:2014dva,Ellis:2014jta,Falkowski:2014tna,Corbett:2015ksa,Buchalla:2015qju,Aad:2015tna,Berthier:2015gja,Falkowski:2015jaa,Englert:2015hrx}. This area, however, is still in its infancy, as such analyses are currently limited
on the experimental side by low statistics.

Top quark physics, on the other hand, has entered a precision era, with data
from the LHC and Tevatron far more abundant. In addition, the top quark plays a
special role in most scenarios of Beyond the Standard Model physics, motivating scrutiny of its phenomenology. Furthermore,
the top sector is strongly coupled to Higgs physics owing to the large top quark
Yukawa coupling, and so represents a complementary window into physics at the electroweak scale. Thus, it is timely to compute the constraints on new top
interactions through a global fit of all dimension-six operators relevant to top
production and decay at hadron colliders.

There have been several studies of the potential for uncovering new physics
effects in the top quark sector at the LHC and Tevatron, phrased in
model-independent language, either through anomalous
couplings~\cite{AguilarSaavedra:2008zc,Bernardo:2014vha,Grzadkowski:2003tf,Nomura:2009tw,Hioki:2009hm,Hioki:2010zu,Hioki:2013hva,Aguilar-Saavedra:2014iga,Chen:2005vr,AguilarSaavedra:2008gt,AguilarSaavedra:2010nx,AguilarSaavedra:2011ct,Fabbrichesi:2014wva,Fabbrichesi:2013bca,Cao:2015doa,GonzalezSprinberg:2011kx}
or higher-dimensional
operators~\cite{Davidson:2015zza,Jung:2014kxa,Cao:2007ea,Degrande:2010kt,Zhang:2010dr,Greiner:2011tt,Degrande:2012wf}. Though
there is a one-to-one correspondence between these two approaches (for the
reasons discussed below) the latter is the approach taken through the rest of
this paper. Other studies have also set limits on top dimension-six operators,
but by considering different physics, such as precision electroweak
data~\cite{deBlas:2015aea}, or flavour-changing neutral
currents~\cite{Aguilar:2015vsa,Durieux:2014xla}.

In a previous work~\cite{Buckley:2015nca}, we published constraints on all
dimension-six operators that contribute to top pair and single top
\textit{production only} in a global fit. Our fitting approach used techniques
borrowed from Monte Carlo event generator tuning, namely the
\textsc{Professor}~\cite{Buckley:2009bj} framework. The purpose of this paper is
to expand on our previous study by adding new measurements, which are sensitive to a
new set of operators not previously examined, including previously
unreleased 8 and 13~\tev data and decay observables, and also to provide a more
detailed review of our general fitting procedure.

The paper is structured as follows. In Section~\ref{sec:ops} we review the
higher-dimensional operators relevant for top quark physics and in
Section~\ref{sec:fit} we review the experimental measurements entering our fit,
as well as the limit-setting procedure we adopt. In Section~\ref{sec:results} we
present our constraints, and discuss the complementarity of LHC and Tevatron
analyses, and the improvements obtained from adding differential distributions
as well as inclusive rates. In Section~\ref{sec:uvmodels} we interpret our
constraints in the context of two specific new physics models. Finally, in
Section~\ref{sec:conclusion} we discuss our results and conclude.

\section{Higher-dimensional operators}
\label{sec:ops}
In effective field theory language, the Standard Model Lagrangian is the first
term in an effective Lagrangian
\begin{equation}
\Leff = \Lsm + \frac{1}{\Lambda}\mathcal{L}_{1}+\frac{1}{\Lambda^{2}}\mathcal{L}_{2}+ \ldots \,,
\end{equation}
\noindent where $\Lambda$ generically represents the scale of the new
physics. From a top-down viewpoint, the higher-dimensional terms that are
suppressed by powers of $1/\Lambda$ originate from heavy degrees of freedom that
have been integrated out. In this way, the low-energy effects of decoupled new
physics can be captured, without the need to consign oneself to a particular
ultraviolet model. The leading contributions to
\Leff at collider energies enter at dimension-six
\begin{equation}
 \Leff = \Lsm + \frac{1}{\Lambda^2}\sum_{i}C_{i}O_{i}(G^a_\mu,W^I_\mu,B_\mu,\varphi,q_L,u_R,d_R,l_L,e_R) +\mathcal{O}(\Lambda^{-4})\,.
\end{equation}
$O_{i}$ are dimension-six operators made up of SM fields, and $C_{i}$ are
dimensionless Wilson coefficients. At dimension-six, assuming minimal flavour
violation and Baryon number conservation, there are 59 independent operators. Clearly, allowing 59 free
parameters to float in a global fit is intractable. Fortunately, for any given
class of observables, only a smaller subset is relevant. In top physics, we
have the following effective operators, expressed in the so-called `Warsaw
basis' of Ref.~\cite{Grzadkowski:2010es}\footnote{Given the simplicity of how it
  captures modifications to SM fermion couplings, this basis is well-suited to
  top EFT. For basis choices of interest in Higgs physics, see
  e.g. Refs.~\cite{Gupta:2014rxa,Giudice:2007fh,Contino:2013kra,Masso:2014xra,Pomarol:2014dya},
  and Ref.~\cite{Falkowski:2015wza} for a tool for translating between them.}
\begin{align}
O^{(1)}_{qq} &= (\bar{q}\gamma_{\mu}q)( \bar{q}\gamma^{\mu}q)  &  O_{uW} &= (\bar{q}\sigma^{\mu \nu} \tau^I u)\tilde \varphi W_{\mu\nu}^{I}  & O^{(3)}_{\varphi q} &= i(\varphi^\dagger \overleftrightarrow{D}^I_\mu \varphi )(\bar{q}\gamma^\mu \tau^I q) \nonumber \\
O^{(3)}_{qq} &= (\bar{q}\gamma_{\mu}\tau^Iq)( \bar{q}\gamma^{\mu}\tau^I q) &  O_{uG} &= (\bar{q}\sigma^{\mu \nu} T^A u)\tilde \varphi G_{\mu\nu}^{A}  & O^{(1)}_{\varphi q} &= i(\varphi^\dagger \overleftrightarrow{D}_\mu \varphi )(\bar{q}\gamma^\mu q)    \nonumber \\
O_{uu} &= (\bar{u}\gamma_{\mu}u)( \bar{u}\gamma^{\mu} u) &  O_{G} &= f_{ABC} G_{\mu}^{A \nu}G_{\nu}^{B \lambda} G_{\lambda}^{C \mu}  &  O_{uB} &= (\bar{q}\sigma^{\mu \nu}u)\tilde \varphi B_{\mu\nu}       \nonumber \\
O^{(8)}_{qu} &=  (\bar{q}\gamma_{\mu}T^Aq)( \bar{u}\gamma^{\mu} T^Au) &  O_{\tilde G} &= f_{ABC} \tilde G_{\mu}^{A \nu}G_{\nu}^{B \lambda} G_{\lambda}^{C \mu}  & O_{\varphi u} &= (\varphi^\dagger i \overleftrightarrow{D}_\mu\varphi)(\bar{u}\gamma^\mu u)    \nonumber \\
O^{(8)}_{qd} &= (\bar{q}\gamma_{\mu}T^Aq)( \bar{d}\gamma^{\mu} T^Ad)  &  O_{\varphi G} &= (\varphi^\dagger \varphi)G_{\mu\nu}^{A}G^{A \mu\nu} \nonumber &  O_{\varphi \tilde G} &= (\varphi^\dagger \varphi) \tilde G_{\mu\nu}^{A}G^{A \mu\nu} \\
O^{(8)}_{ud} &= (\bar{u}\gamma_{\mu}T^Au)( \bar{d}\gamma^{\mu} T^Ad) \,.
\label{eqn:allops}
\end{align}
%
We adopt the same notation as Ref.~\cite{Grzadkowski:2010es}, where $T^A=\tfrac{1}{2}\lambda^A$ are the $SU(3)$ generators, and $\tau^{I}$ are the Pauli matrices, related to the generators of $SU(2)$ by $S^I=\tfrac{1}{2}\tau^I$. For the four-quark operators on the left column of eq. (\ref{eqn:allops}), we denote a specific flavour combination $(\bar{q}_i...q_j)(\bar{q}_k...q_l)$ by e.g. $O_{4q}^{ijkl}$. It should be noted that the operators $O_{uW}$, $O_{uG}$ and $O_{uB}$ are not hermitian and so may
have complex coefficients which, along with $O_{\tilde G}$ and
$O_{\varphi \tilde G}$, lead to $\mathcal{CP}$-violating effects. These do not
contribute to Standard Model spin-averaged cross-sections, though they are in
principle sensitive to polarimetric observables such as spin correlations, and
should therefore be treated as independent operators. However, currently available
measurements that would be sensitive to these degrees of freedom have been extracted by making model-specific assumptions that preclude their usage in our fit, e.g. by assuming that the tops are produced with either SM-like spin correlation or no spin correlation at all, as in Refs.~\cite{Chatrchyan:2013wua,Aad:2014pwa}. We will discuss this issue in more detail in the next section. With these caveats, a
total of 14 constrainable $\mc{CP}$-even dimension-six operators contribute to top quark
production and decay at leading order in the SMEFT.

\section{Methodology}
\label{sec:fit}

\subsection{Experimental inputs}
The experimental measurements used in the
fit~\cite{Aad:2014kva,ATLAS:2012aa,Aad:2012mza,Aad:2012qf,Aad:2012vip,Aad:2014jra,Aad:2015pga,Chatrchyan:2013ual,Chatrchyan:2012bra,Chatrchyan:2012ria,Chatrchyan:2012vs,Chatrchyan:2013kff,Chatrchyan:2013faa,Khachatryan:2015uqb,Aaltonen:2013wca,Aad:2014fwa,Aaltonen:2014qja,Khachatryan:2014iya,Abazov:2009pa,Abazov:2011rz,Aad:2014zka,Aaltonen:2009iz,Chatrchyan:2012saa,Khachatryan:2015oqa,Abazov:2014vga,Aad:2013cea,Chatrchyan:2014yta,Aaltonen:2012it,Abazov:2014cca,Aaltonen:2013kna,Abazov:2012vd,Aad:2012ky,Aaltonen:2012lua,Chatrchyan:2013jna,Abazov:2010jn,Aad:2015uwa,Aad:2015eua,Khachatryan:2014ewa}
are included in Table~\ref{table:measurements}. All these measurements are
quoted in terms of `parton-level' quantities; that is, top quarks and their
direct decay products. Whilst it is possible to include particle-level
observables, these are far less abundant and they are beyond the scope of the
present study.

The importance of including kinematic distributions is manifest here. For top
pair production, for instance, we have a total of 195 measurements, 174 of which
come from differential observables. This size of fit is unprecedented in top
physics, which underlines the need for a systematic fitting approach, as
provided by \textsc{Professor}. Indeed top pair production cross-sections make up
the bulk of measurements that are used in the fit. Single top production
cross-sections comprise the next dominant contribution. We also make
use of data from charge asymmetries in top pair production, as well as inclusive measurements of top pair production in association with a photon or a $Z$ ($t\bar{t}\gamma$ and $t\bar{t}Z$) and
observables relating to top quark decay. We take each of these categories of
measurement in turn, discussing which operators are relevant and the constraints
obtained on them from data.

\begin{table*}[!t]
\begin{center}
\begin{adjustbox}{max width=\textwidth}
\begin{tabular}{l r  l  l || l  r  l  l  }
\toprule
Dataset & $\sqrt{s}$ (\tev) & {Measurements} & {arXiv ref.} & Dataset & $\sqrt{s}$ (\tev) & {Measurements} & {arXiv ref.} \\
\midrule
\multicolumn{8}{l}{\textit{Top pair production}} \\
\multicolumn{4}{l}{Total cross-sections:} & \multicolumn{4}{l}{Differential cross-sections:} \\
ATLAS & 7  & lepton+jets   & 1406.5375 & ATLAS & 7  &  $p_T (t),M_{t\bar{t}},|y_{t\bar{t}}|$  & 1407.0371   \\
ATLAS & 7  & dilepton   &  1202.4892 & CDF    & 1.96   & $M_{t\bar{t}}$  & 0903.2850  \\
ATLAS & 7  & lepton+tau   & 1205.3067 & CMS    & 7      &  $p_T (t),M_{t\bar{t}},y_t,y_{t\bar{t}} $  &  1211.2220  \\
ATLAS & 7  & lepton w/o $b$ jets   & 1201.1889 & CMS    & 8      &  $p_T (t),M_{t\bar{t}},y_t,y_{t\bar{t}}$  & 1505.04480  \\
ATLAS & 7  & lepton w/ $b$ jets   & 1406.5375 & \dzero      & 1.96  & $M_{t\bar{t}},p_T(t),|y_t|$ &  1401.5785  \\
ATLAS & 7  & tau+jets   & 1211.7205 & & & \\
ATLAS & 7  & $t\bar{t},Z\gamma,WW$   & 1407.0573 & \multicolumn{4}{l}{Charge asymmetries:}  \\
ATLAS & 8  & dilepton  & 1202.4892 & ATLAS & 7  &  \Ac (inclusive+$M_{t\bar{t}},y_{t\bar{t}}$)  & 1311.6742  \\
CMS & 7  & all hadronic   & 1302.0508 & CMS & 7  &  \Ac (inclusive+$M_{t\bar{t}},y_{t\bar{t}}$)  & 1402.3803 \\
CMS & 7  & dilepton  & 1208.2761 & CDF & 1.96 & \Afb   (inclusive+$M_{t\bar{t}},y_{t\bar{t}}$) & 1211.1003  \\
CMS & 7  & lepton+jets   & 1212.6682 &  \dzero & 1.96 & \Afb  (inclusive+$M_{t\bar{t}},y_{t\bar{t}}$) & 1405.0421 \\
CMS & 7  & lepton+tau   & 1203.6810 &  & & \\
CMS & 7  & tau+jets   & 1301.5755 & Top widths: &  & \\
CMS & 8  & dilepton  &  1312.7582 & \dzero & 1.96 & $\Gamma_{\!\mathrm{top}}$& 1308.4050 \\
CDF + \dzero & 1.96  & Combined world average   & 1309.7570  & CDF & 1.96 & $\Gamma_{\!\mathrm{top}}$ & 1201.4156  \\
\addlinespace
\multicolumn{4}{l}{\textit{Single top production}} & \multicolumn{4}{l}{ $W$\!-boson helicity fractions:} \\
ATLAS & 7  & $t$-channel (differential)   & 1406.7844 & ATLAS & 7  & & 1205.2484  \\
CDF & 1.96  & $s$-channel (total)   & 1402.0484 & CDF & 1.96 & & 1211.4523  \\
CMS & 7  & $t$-channel (total)   & 1406.7844 & CMS  & 7  & & 1308.3879  \\
CMS & 8  & $t$-channel (total)   & 1406.7844 & \dzero & 1.96  & & 1011.6549   \\
\dzero & 1.96  & $s$-channel (total)   & 0907.4259 & & & \\
\dzero & 1.96  & $t$-channel (total)   & 1105.2788 & & & \\
\addlinespace
\multicolumn{4}{l}{\textit{Associated production}} & \multicolumn{4}{l}{\textit{Run~II data}}\\
ATLAS & 7  & $t\bar{t}\gamma$   & 1502.00586 & CMS & 13 & $t\bar{t}$ (dilepton) & 1510.05302 \\
ATLAS & 8  & $t\bar{t}Z$   & 1509.05276 &&&& \\
CMS & 8  & $t\bar{t}Z$   & 1406.7830 &&&& \\
\bottomrule
\end{tabular}
\hspace{0.2cm}
\end{adjustbox}
\end{center}
\caption{\label{table:measurements}
 The measurements entering our fit. Details of each are described in the text.}
\end{table*}

\subsection{Treatment of uncertainties}
The uncertainties entering our fit can be classed into three categories:

\begin{cedescription}
\item[Experimental uncertainties:]
  We generally have no control over these. In cases where statistical and
  systematic (and luminosity) errors are recorded separately, we add them in
  quadrature. Correlations between measurements are also an issue: the unfolding
  of measured distributions to parton-level introduces some correlation
  between neighbouring bins. If estimates of these effects have been provided in
  the experimental analysis, we use this information in the fit, if they are
  not, we assume zero correlation. However, we have checked that bin
  correlations have little effect on our numerical results.

  There will also be correlations between apparently separate measurements. The
  multitude of different top pair production cross-section measurements will
  clearly be correlated due to overlapping event selection criteria and detector
  effects, etc. Without a full study of the correlations between different decay
  channels measured by the same experiment, these effects cannot be completely
  taken into account, but based on the negligible effects of the bin-by-bin correlations on
  our numerical results we can expect these effects to be small as well.

\item[Standard Model theoretical uncertainties:]
 These stem from the choice of parton distribution functions (PDFs), as well as
  neglected higher-order perturbative corrections. As is conventional, we model
  the latter by varying the renormalisation and factorisation scales
  independently in the range $\mu_0/2\leq\mu_\mathrm{R,F}\leq 2\mu_0$, where
  we use $\mu_0=m_t$ as the default scale, and take the envelope as our
  uncertainty. For the PDF uncertainty, we follow the PDF4LHC
  recommendation~\cite{Butterworth:2015oua} of using
  CT10~\cite{Nadolsky:2008zw}, MSTW~\cite{Martin:2009iq} \&
  NNPDF~\cite{Ball:2010de} NLO fits, each with associated scale uncertainties,
  then taking the full width of the scale+PDF envelope as our uncertainty
  estimate -- i.e. we conservatively assume that scales and parton densities are
  100\% correlated. Unless otherwise stated, we take the top quark mass to be
  $m_t = 173.2 \pm 1.0~\gev$. We do not consider electroweak corrections.

  Only recently a lot of progress has been made in extending the dimension
  six-extended SM to higher order, see
  Refs.~\cite{Passarino:2012cb,Mebane:2013zga,Jenkins:2013zja,Jenkins:2013sda,Jenkins:2013wua,Alonso:2013hga,Hartmann:2015oia,Ghezzi:2015vva,Zhang:2013xya,Englert:2014cva,Hartmann:2015aia,Cheung:2015aba,Drozd:2015rsp,Gauld:2015lmb}. Including
  these effects is beyond the scope of this work, also because we work to
  leading order accuracy in the electroweak expansion of the SM. QCD corrections to four fermion operators included via renormalisation group equations are typically of the order of 15\%, depending on the resolved phase space~\cite{Englert:2014cva}.  As pointed out
  in Ref.~\cite{Berthier:2015oma}, these effects can be important in
  electroweak precision data fits.

\item[Interpolation error:]
  A small error relating to the Monte Carlo interpolation (described in more
  detail in the next section) is included. This is estimated to be 5\% at a
  conservative estimate, as discussed in the following section, and subleading compared to the previous two categories.

\end{cedescription}

\subsection{Fitting procedure}

Our fitting procedure, briefly outlined in Ref.~\cite{Buckley:2015nca}, uses the
\textsc{Professor} framework. The first step is to construct an $N$-dimensional
hypercube in the space of dimension six couplings, compute the
observables at each point in the space, and then to fit an {\it interpolating
  function} $f(\mathbf{C})$ that parametrises the theory prediction as a
function of the Wilson coefficients $\mathbf{C}=\{C_i\}$. This can then be used
to rapidly generate theory observables for arbitrary values of the
coefficients. Motivated by the dependence of the total cross-section with a
Wilson coefficient:
\begin{equation}
\sigma \sim \sigma_\mathrm{SM} + C_i\sigma_{D6}+ C^2_i\sigma_{D6^2}\, ,
\label{eqn:sigmad6}
\end{equation}
\noindent the fitting function is chosen to be a second-order or higher polynomial:
\begin{equation}
f_b(\{C_i\}) = \alpha_0^b + \sum_{\substack{i}} \beta_i^bC_i  + \sum_{\substack{i\leq j}}\gamma^b_{i,j}C_iC_j+\ldots\,.
\label{eqn:ipol}
\end{equation}

\begin{figure}[!t]
\begin{center}
\includegraphics[width=0.45\textwidth]{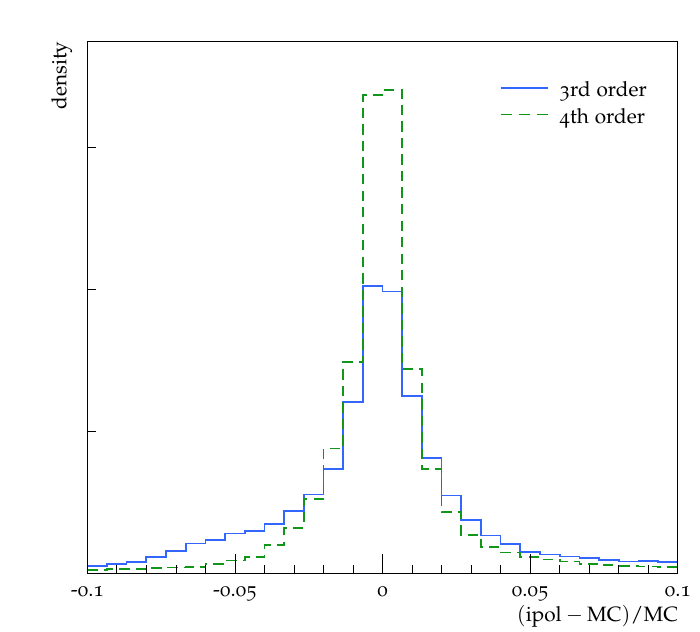}
\quad
\includegraphics[width=0.45\textwidth]{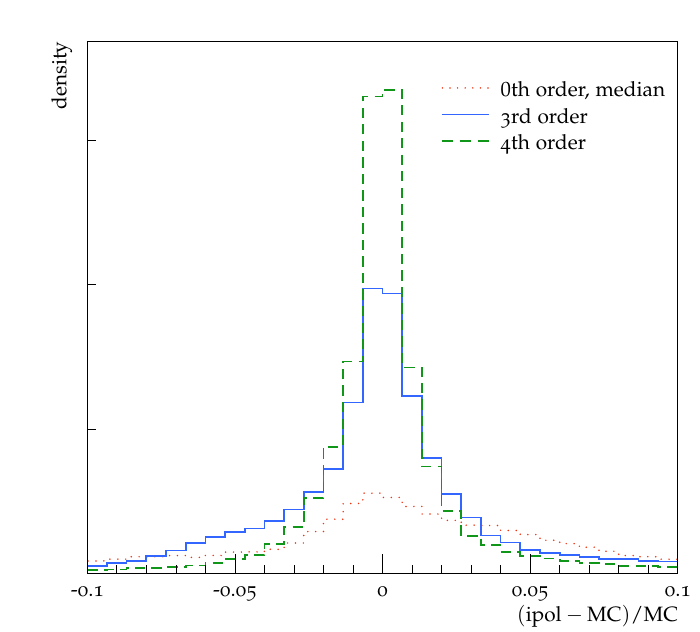}
\caption{Residuals distributions for interpolated observable values~(left) and
  uncertainties~(right), evaluated over all input MC runs and all observables. The 4th
  order polynomial parameterisation gives the best performance and the vast
  majority of entries are within 5\% of the explicit MC value. The poor
  performance of a constant uncertainty assumption based on the median input
  uncertainty is evident -- since all three lines have the same normalisation, the
  majority of residual mismodellings for the median approach are (far) outside
  the displayed 10\% interval.}
\label{fig:residuals}
\end{center}
\end{figure}

In the absence of systematic uncertainties, each observable would exactly follow
a second-order polynomial in the coefficients, and higher-order terms capture
bin uncertainties which modify this. The polynomial also serves as a useful
check that the dimension-six approximation is valid. By comparing
eq.~(\ref{eqn:sigmad6}) with eq.~(\ref{eqn:ipol}), we see that the terms
quadratic in $C_i$ are small provided that the coefficients in the interpolating
function $\gamma_{i,j}$ are small. This is a more robust way to ensure validity
of the dimension-six approximation than to assume a linear fit from the
start.

In practice, to minimise the interpolation uncertainty, we use up to a 4th order
polynomial in eq.~(\ref{eqn:ipol}), depending on the observable of interest. The
performance of the interpolation method is shown in Figure~\ref{fig:residuals},
which depicts the fractional deviation of the polynomial fit from the explicit
MC points used to constrain it. The central values and the sizes of the
modelling uncertainties may both be parameterised with extremely similar
performance, with 4th order performing best for both. The width of this residual
mismodeling distribution being $\sim \text{3\%}$ for each of the value and
error components is the motivation for a total 5\% interpolation uncertainty to
be included in the goodness of fit of the interpolated MC polynomial $f(\mathbf{C})$ to the experimentally measured value $E$:
\begin{equation}
  \chi^2(\mathbf{C}) = \sum_{\substack{\mathcal{O}} } \sum_{\substack{i,j} }\frac{(f_i(\mathbf{C}) - E_i)\rho_{i,j} (f_j(\mathbf{C}) - E_j)}{\sigma_i\sigma_j} \,,
\end{equation}
where we sum over all observables $\mathcal{O}$ and all bins in that observable
$i$. We include the correlation matrix $\rho_{i,j}$ where this is provided by
the experiments, otherwise $\rho_{i,j}=\delta_{ij}$. The uncertainty on each bin
is given by $\sigma_i = \sqrt{\sigma_{\mathrm{th},i}^2+\sigma_{\mathrm{exp},i}^2}$, i.e. we treat
theory and experimental errors as uncorrelated. The
parameterisation of the theory uncertainties is restricted to not become larger
than in the training set, to ensure that polynomial blow-up of the uncertainty
at the edges of the sampling range cannot produce a spuriously low $\chi^2$ and
disrupt the fit.

We hence have constructed a fast parameterisation of model goodness-of-fit as a
function of the EFT operator coefficients. This may be used to produce $\chi^2$
maps in slices or marginalised projections of the operator space, which are then
transformed to confidence intervals on the coefficients $C_i$, defined by the
regions for which
\begin{equation}
 1 - \mathrm{CL} \geq \int^{\infty}_{\chi^2(C_i)} f_k(x) dx \,,
\end{equation}
where typically $\mathrm{CL} \in \{0.68,0.95,0.99\}$ and $f_k(x)$ is the
$\chi^2$ distribution for $k$ degrees of freedom, which we define as
$k = N_\mathrm{measurements} - N_\mathrm{coefficients}$.

\section{Results}
\label{sec:results}

The entire 59 dimensional operator set of
Ref.~\cite{Grzadkowski:2010es} was implemented in a
FeynRules~\cite{Christensen:2008py} model file. The contributions to
parton level cross-sections and decay observables from the above
operators were computed using
\textsc{MadGraph/Madevent}~\cite{Alwall:2014hca}, making use of the
Universal FeynRules Output (UFO)~\cite{Degrande:2011ua} format. We
model NLO~QCD corrections by including Standard Model $K$-factors
(bin-by-bin for differential observables), where the NLO observables
are calculated using MCFM~\cite{Campbell:2010ff}, cross-checked with
MC@NLO~\cite{Frixione:2002ik,Frixione:2010wd}. These $K$-factors are used for arbitrary
values of the Wilson coefficients, thus modelling NLO effects in the
pure-SM contribution only.  More specifically, this amounts to
performing a simultaneous expansion of each observable in the strong
coupling $\alpha_s$ and the (inverse) new physics scale
$\Lambda^{-1}$, and neglecting terms $\sim{\cal
  O}(\alpha_S\Lambda^{-2})$. Our final 95\% confidence limits for
each coefficient are presented in Figure~\ref{fig:constraints}; we
discuss them in more detail below.

\subsection{Top pair production}
\label{sec:toppair}
By far the most abundant source of data in top physics is from the production of
top pairs. The $\mc{CP}$-even dimension-six operators that interfere with the
Standard Model amplitude are 
\begin{equation}
\begin{split}
\mc{L}_{D6} & \supset  \frac{C_{uG}}{\Lambda^2} (\bar{q}\sigma^{\mu \nu} T^A u)\tilde \varphi G_{\mu\nu}^{A} +  \frac{C_{G}}{\Lambda^2}  f_{ABC} G_{\mu}^{A \nu}G_{\nu}^{B \lambda} G_{\lambda}^{C \mu} + \frac{C_{\varphi G}}{\Lambda^2} (\varphi^\dagger \varphi)G_{\mu\nu}^{A}G^{A \mu\nu} \\
& + \frac{C^{(1)}_{qq}}{\Lambda^2}(\bar{q}\gamma_{\mu}q)( \bar{q}\gamma^{\mu}q) +  \frac{C^{(3)}_{qq}}{\Lambda^2}(\bar{q}\gamma_{\mu}\tau^Iq)( \bar{q}\gamma^{\mu}\tau^I q) +  \frac{C_{uu}}{\Lambda^2}(\bar{u}\gamma_{\mu}u)( \bar{u}\gamma^{\mu} u) \\
& + \frac{C^{(8)}_{qu}}{\Lambda^2}(\bar{q}\gamma_{\mu}T^Aq)( \bar{u}\gamma^{\mu} T^Au) +  \frac{C^{(8)}_{qd}}{\Lambda^2}(\bar{q}\gamma_{\mu}T^Aq)( \bar{d}\gamma^{\mu} T^Ad) +  \frac{C^{(8)}_{ud}}{\Lambda^2}(\bar{u}\gamma_{\mu}T^Au)( \bar{d}\gamma^{\mu} T^Ad) \,.
\end{split}
\label{eqn:ttbarops}
\end{equation}

As pointed out in Ref.~\cite{Buckley:2015nca}, the operator $O_{\varphi G}$ cannot be
bounded by top pair production alone, since the branching ratio to virtual top
pairs for a 125~\gev Higgs is practically zero, therefore we do not consider it
here. For a recent constraint from Higgs physics see
e.g. Ref.~\cite{Corbett:2015ksa,Ellis:2014jta,Falkowski:2015jaa,Englert:2015hrx}. We further ignore the contribution of the operator $O^{11}_{uG}$\,, as this operator is a direct mixing of the left- and right- chiral $u$ quark fields, and so contributes terms proportional to $m_u$. We
also note that the six four-quark operators of eq.
(\ref{eqn:ttbarops}) interfere with the Standard Model QCD processes $\bar{u}u,\,\bar{d}d\,\rightarrow\,\bar{t}t$ to produce terms dependent only on the four linear
combinations of Wilson Coefficients (following the notation of Ref.~\cite{Zhang:2010dr})
\begin{equation}
\begin{split}
 C^1_u = &~C^{(1) {1331}}_{qq}+ C^{1331}_{uu}+ C^{(3) {1331}}_{qq} \\
 C^2_u = &~C^{(8) {1133}}_{qu} +  C^{(8) {3311}}_{qu} \\
 C^1_d = &~C^{(3) {1133}}_{qq}+\tfrac{1}{4}C^{(8) {3311}}_{ud} \\
 C^2_d = &~C^{(8) {1133}}_{qu} +  C^{(8) {3311}}_{qd}\,.
\end{split}
\label{eqn:4fs}
\end{equation}

\begin{figure}[!t]
\begin{center}
\includegraphics[width=0.95\textwidth]{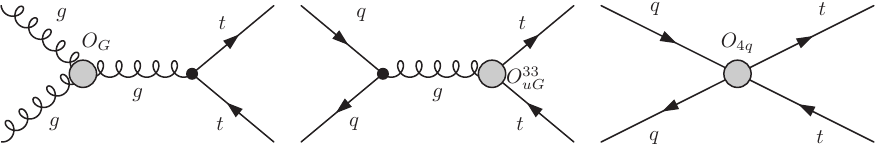}
\caption{Sample Feynman diagrams for the interference of the leading-order SM amplitudes for top pair production with the operators of eq. (\ref{eqn:ttbarops}). $O_{4q}$ denotes the insertion of any of the four-quark operators. }
\label{fig:ttbarfeyn}
\end{center}
\end{figure}

It is these four that are constrainable in a dimension-six analysis. Finally, we
note that the operator $O_G$, whilst not directly coupling to the top at tree-level, should not be neglected. Since it modifies the triple gluon vertex, and the $gg$ channel contributes $\sim 75\%$ $(90\%)$ of the total top pair production cross-section at the 8 (13) \tev LHC, moderate values of its Wilson
coefficient can substantially impact total rates. We note, however, that in this special case, the cross section modifications are driven by the squared dimension six terms instead of the linearised interference with the SM. Nonetheless, in the interests of generality, we choose to include this operator in our fit at this stage, noting that bounds on its Wilson coefficient should be interpreted with caution.\footnote{We have observed that excluding this operator actually tightens the bounds on the remaining ones, so choosing to keep it is the more conservative option.} Representative Feynman diagrams for the interference of these operators are
shown in Figure~\ref{fig:ttbarfeyn}.

\begin{figure}[!t]
\begin{center}
\includegraphics[width=\textwidth]{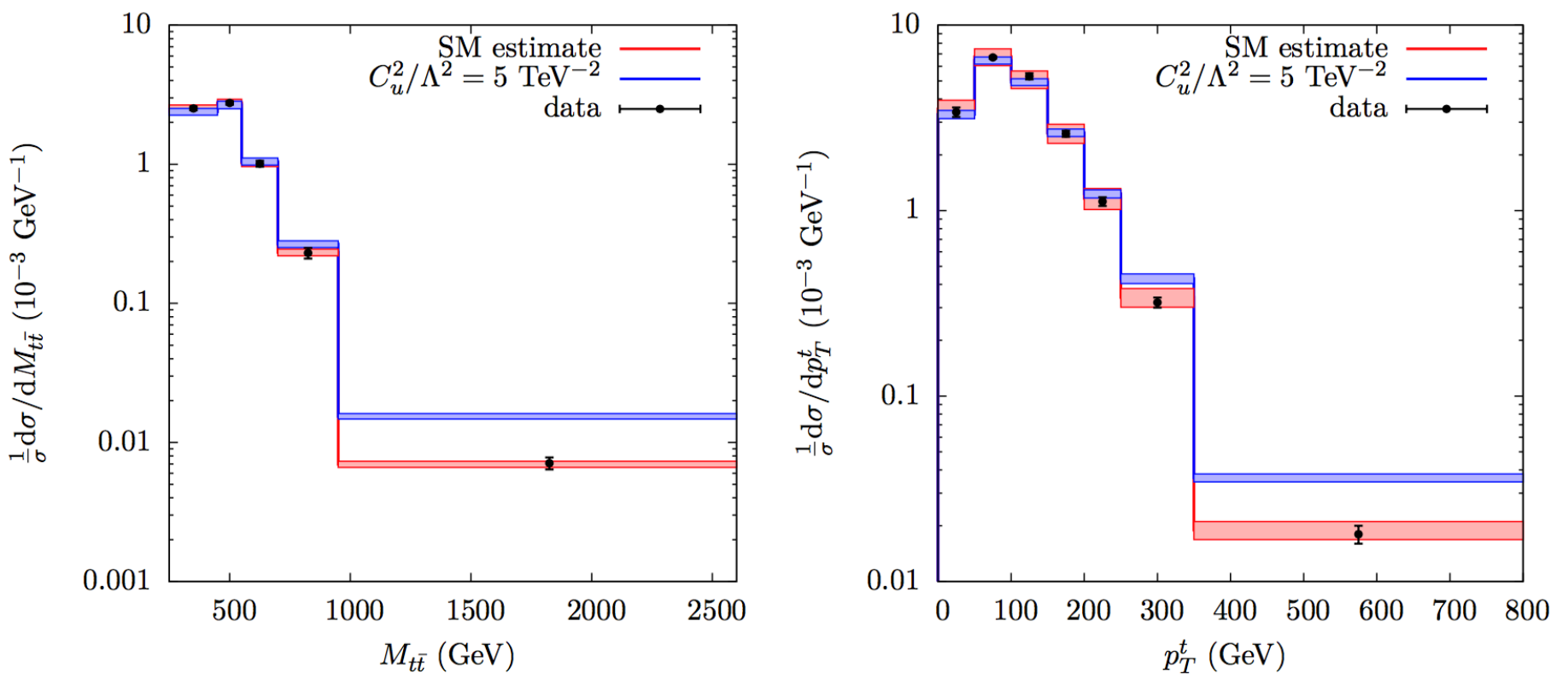}
\caption{Parton level differential distributions in top pair production, considering SM only (red) and the effects of the four-quark operator $O^2_u$, showing the enhancement in the tails of the distributions. Data taken from Ref.~\cite{Aad:2014zka}.}
\label{fig:distributions}
\end{center}
\end{figure}

The most obvious place to look for the effects of higher-dimensional terms is
through the enhancement (or reduction, in the case of destructive interference)
of total cross-sections. Important differences between SM and dimension-six
terms are lost in this approach, however, since operators can cause
deviations in the shape of distributions without substantially impacting event
yields. This is highlighted in Figure~\ref{fig:distributions}, where we plot our
NLO SM estimate for two top pair distributions, vs.  one with a large
interference term. Both are consistent with the data in the threshold region,
which dominates the cross-section, but clear discrimination between SM and
dimension-six effects is visible in the high-mass region, which simply originates from the scaling of dimension-six operator effects as $s/\Lambda^2$\footnote{One may worry that the inclusion of the final `overflow' bin in the invariant mass distributions may invalidate the EFT approach. We have performed the global fit without these data points, and found that they have little effect on our constraints. This is due to the large experimental uncertainties in this region, and the fact that these bins comprise less than 5\% of the total degrees of freedom in our fit, so have little statistical pull.} .

\begin{figure}[!t]
\begin{center}
\includegraphics[width=\textwidth]{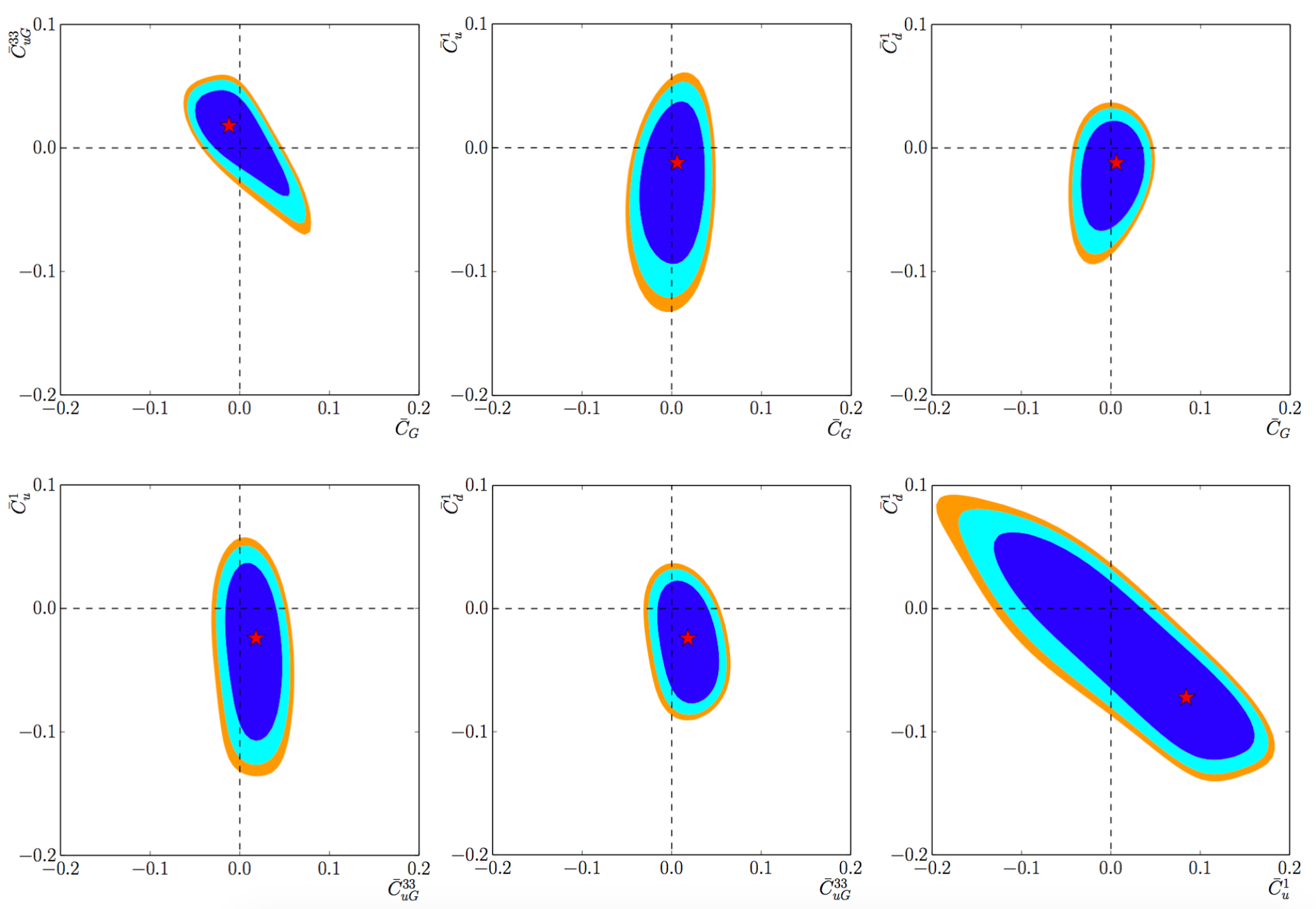}
\caption{68\%, 95\% and 99\% confidence intervals for selected combinations of operators contributing to top pair production, with all remaining operators set to zero. The star marks the best fit point, indicating good agreement with the Standard Model. Here $\bar{C}_i = C_iv^2/\Lambda^2$.}
\label{fig:toppairops}
\end{center}
\end{figure}

\begin{figure}[!t]
\begin{center}
\includegraphics[width=\textwidth]{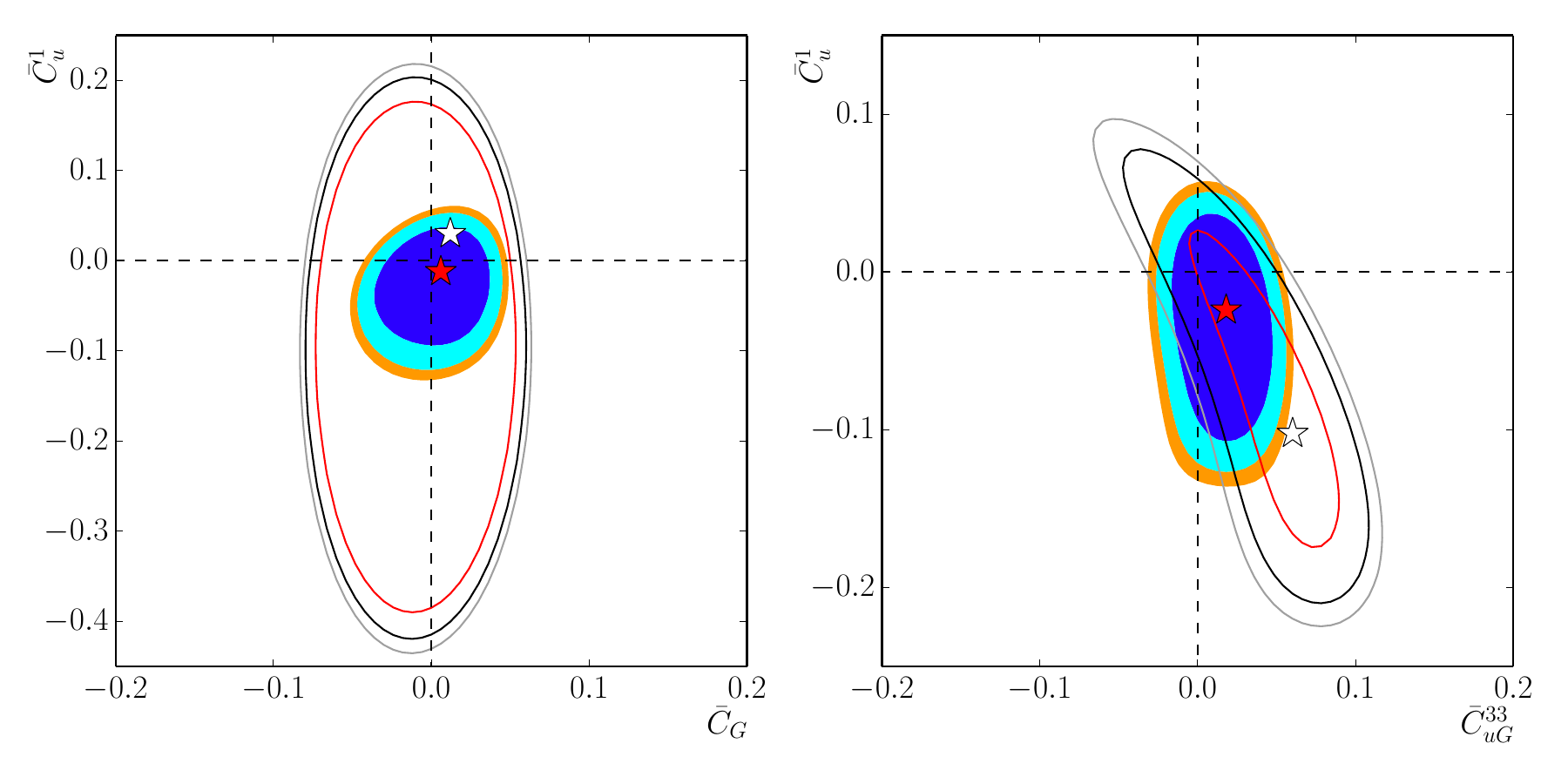}
\caption{Left: 68\%, 95\% and 99\% confidence intervals on the operators $C_G$ vs. $C^1_u$ , considering differential and total cross-sections (contours, red star), and total cross-sections only (lines, white star). Right: Limits on $C^{33}_{uG}$ vs. $C^1_u$, considering both Tevatron and LHC data (contours) and Tevatron data only (lines).}
\label{fig:c3c4diff}
\end{center}
\end{figure}

Limits on these operators can be obtained in two ways; by setting all other
operators to zero, and by marginalising over the other parameters in a global
fit. In Figure~\ref{fig:toppairops} we plot the
allowed 68\%, 95\% and 99\% confidence intervals for various pairs of operators, with all others set to zero,
showing correlations between some coefficients. Most of these operators appear
uncorrelated, though there is a strong correlation between $C^1_u$ and $C^1_d$,
due to a relative sign between their interference terms. Given the
lack of reported deviations in top quark measurements, it is perhaps unsurprising to
see that all Wilson coefficients are consistent with zero within the 95\% confidence intervals, and that
the SM hypothesis is an excellent description of the data. In Figure~\ref{fig:c3c4diff}, the stronger joint constraints on $C_G$ vs $C^1_u$ obtained from including differential measurements make manifest the importance of utilizing all available cross-section information.

It is also interesting to note the relative pull of measurements from the LHC
and Tevatron, as illustrated in Figure~\ref{fig:c3c4diff}. It is interesting to see
that although Tevatron data are naively more sensitive to
four-quark operators, after the LHC Run~I and early into Run~II, the LHC data size and probed energy transfers lead to comparably stronger constraints. In our fit this is highlighted by the simple fact that LHC data comprise
more than 80\% of the bins in our fit, so have a much larger pull. This
stresses the importance of collecting large statistics as well as using
sensitive discriminating observables.

\subsection{Single top production}
The next most abundant source of top quark data is from single top
production. In our fit we consider production in the $t$ and $s$
channels, and omit $Wt$-associated production. Though measurements of
the latter process have been published, they are not suitable for
inclusion in a fit involving parton level theory predictions. As is
well-known, $Wt$ production interferes with top pair production at NLO
and beyond in a five-flavour
scheme~\cite{Zhu:2001hw,Campbell:2005bb,Cao:2008af}, or at LO in a
four-flavour one. Its separation from top pair production is then a
delicate issue, discussed in detail in
Refs.~\cite{Frixione:2008yi,White:2009yt,Kauer:2001sp,Kersevan:2006fq}. We
thus choose to postpone the inclusion of $Wt$ production to a future
study, going beyond parton level. The operators that could lead to
deviations from SM predictions are shown below
\begin{equation}
\begin{split}
\mc{L}_{D6}	& \supset  \frac{C_{uW}}{\Lambda^2} (\bar{q}\sigma^{\mu \nu} \tau^I u)\,\tilde \varphi\, W_{\mu\nu}^{I} + \frac{C^{(3)}_{\varphi q}}{\Lambda^2} i(\varphi^\dagger \overleftrightarrow{D}^I_\mu \varphi )(\bar{q}\gamma^\mu \tau^I q) \\
			& + \frac{C_{\varphi ud}}{\Lambda^2} (\varphi^\dagger  \overleftrightarrow{D}_\mu \varphi )(\bar{u}\gamma^\mu d) +  \frac{C_{dW}}{\Lambda^2} (\bar{q}\sigma^{\mu \nu} \tau^I d)\,\tilde \varphi \,W_{\mu\nu}^{I}  \\
			& + \frac{C^{(3)}_{qq}}{\Lambda^2}(\bar{q}\gamma_{\mu}\tau^Iq)( \bar{q}\gamma^{\mu}\tau^I q) + \frac{C^{(1)}_{qq}}{\Lambda^2}(\bar{q}\gamma_{\mu}q)( \bar{q}\gamma^{\mu}q)  + \frac{C^{(1)}_{qu}}{\Lambda^2}(\bar{q}\gamma_{\mu}q)( \bar{u}\gamma^{\mu} u)\,.
\end{split}
\label{eqn:singletopops}
\end{equation}

As in top~pair production there are several simplifications which reduce
this operator set. The right-chiral down quark fields appearing in $O_{dW}$ and $O_{\varphi ud}$ cause these operators' interference with the left-chiral SM weak interaction to be proportional to the relevant down-type quark mass. For example, an operator insertion of $O^{33}_{\varphi ud}$ will always contract with the SM $Wtb$\,-vertex to form a term of order $m_b\, m_t\, C^{33}_{\varphi ud}/\Lambda^2$. Since $m_b$ is much less than both $\hat{s}$ and the other dimensionful parameters that appear, $v$ and $m_t$, we may choose to neglect these operators. By the same rationale we neglect $O^{(1)}_{qu}$ as its contribution to observables is $\mathcal{O}(m_u)$. We have further checked numerically that the contribution of these operators is practically negligible.  Finally, all contributing four-fermion partonic subprocesses depend only on the linear combination of Wilson Coefficients:

\begin{equation}
\begin{split}
 C_t = &~C^{(3) {1331}}_{qq} + \tfrac{1}{6}(C^{(1) {1331}}_{qq}- C^{(3) 1331}_{qq}).
\end{split}
\label{eqn:4fs2}
\end{equation}

Single top production can thus be characterised by the three dimension-six operators $O_{uW}$, $O^{(3)}_{\varphi q}$ and $O_t$.

As noted in the introduction, several model-independent studies have noted the
potential for uncovering new physics in single top production, though these have
typically been expressed in terms of anomalous couplings, via the Lagrangian
\begin{equation}
\mathcal{L}_{Wtb} = \frac{g}{\sqrt{2}} \bar{b} \gamma^\mu (V_L P_L + V_R P_R) t W^{-}_{\mu} +  \frac{g}{\sqrt{2}}\bar{b}\frac{i\sigma^{\mu\nu}q_\nu}{M_W}(g_LP_L + g_RP_R)tW^-_\mu + h.c. \,
\end{equation}
where $q = p_t - p_b$. There is a one-to-one mapping between this Lagrangian and those dimension-six
operators that modify the $Wtb$ vertex:
%
\begin{align}
  V_L & \to V_{tb} + C^{(3)}_{\varphi q}v^2/\Lambda^2 & V_R & \to \frac{1}{2}C_{\varphi u d}v^2/\Lambda^2 \nonumber \\
  g_L &  \to \sqrt{2}C_{uW}v^2/\Lambda^2 & g_R &  \to \sqrt{2}C_{dW}v^2/\Lambda^2
\end{align}

\begin{figure}[!t]
\begin{center}
\includegraphics[width=\textwidth]{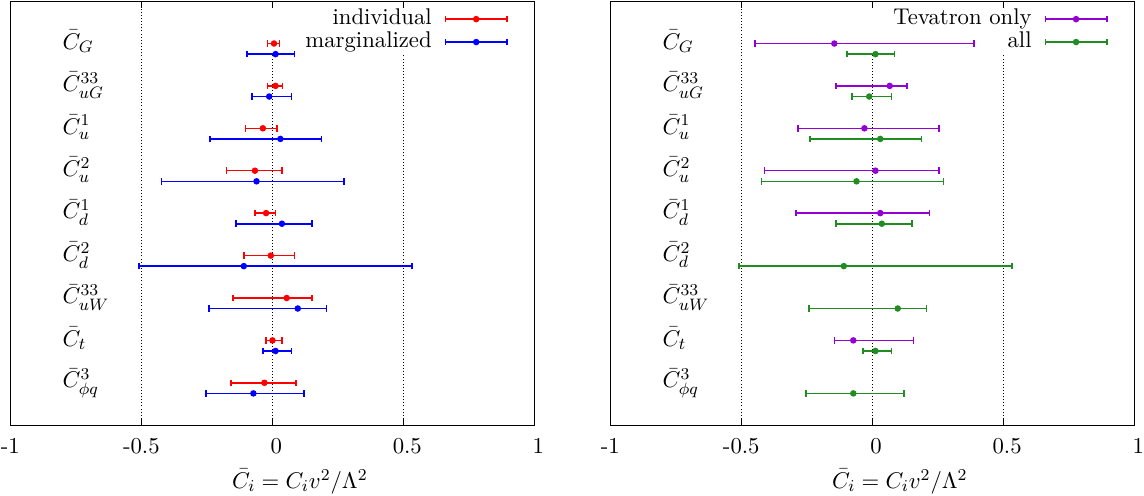}
\caption{Left: Individual (red) and marginalised (blue) 95\% confidence intervals on dimension-six operators from top pair production and single top production (bottom three). Right: Marginalised 95 \% bounds considering all data from LHC and Tevatron (green) vs Tevatron only (purple).}
\label{fig:allvstev}
\end{center}
\end{figure}

What, then, is the advantage of using higher-dimensional operators when
anomalous couplings capture most of the same physics? The advantages are
manifold. Firstly, the power-counting arguments of the previous paragraph that
allowed us to reject the operators $O_{dW}$, $O_{\varphi ud}$ at order
$\Lambda^{-2}$ would not be clear in an anomalous coupling framework. In
addition, the four-quark operator $O^{(3)}_{qq}$ in eq. (\ref{eqn:singletopops}) can
have a substantial effect on single-top production, but this can only be
captured by an EFT approach. For a detailed comparison of these approaches, see e.g.~Ref.~\cite{Zhang:2010px}. The 95\% confidence limits on these operators from
single top production are shown in Fig. (\ref{fig:allvstev}), along with those operators previously
discussed in top pair production.

Let us compare these results to our findings of Section~\ref{sec:toppair}. The bounds on operators from top pair production are typically stronger. The so-called chromomagnetic moment operator $O_{uG}$ is also tightly constrained, owing to its appearance in both the $q\bar{q}$ and $gg$ channels,
i.e. it is sensitive to both Tevatron and LHC measurements.  For the four-quark
operators, the stronger bounds are typically on the $C^1_i$-type. This
originates from the more pronounced effect on kinematic distributions that they
have. The phenomenology of the $C^2_i$-type operators is SM-like, and their effect becomes only visible in the tails of distributions.

The much wider marginalised bounds on these two operators stems
from the relative sign between their interference term and those of the other
operators, which results in cancellations in the total cross-section that
significantly widen the allowed ranges of $C_i$. With the exception of $C_t$, which strongly modifies the single top production cross-section, the individual bounds on the operator coefficients from single top
production are typically weaker. This originates from the larger experimental
uncertainties on single top production, that stem from the multitude of
different backgrounds that contaminate this process, particularly top pair
production. For the Tevatron datasets this is particularly telling: the few
measurements that have been made, with no differential distributions, combined
with the large error bars on the available data, mean that two of the three operators are
not constrained at dimension-six\footnote{Our bounds on these two operators are
  of the same order, but wider, than a pre-LHC phenomenological study~\cite{Cao:2007ea}, owing to
  larger experimental errors than estimated there.}. Still, as before, excellent
agreement with the SM is observed.

In addition to single-top production, the operator $O_{uW}$ may be constrained
by distributions relating to the kinematics of the top quark decay. The matrix
element for hadronic top quark decay $t\to Wb \to b q q'$, for instance, is
equivalent to that for $t$-channel single top production via crossing symmetry,
so decay observables provide complementary information on this operator. We will
discuss the bounds obtainable from decay observables in Section~\ref{sec:decays}.

\subsection{Associated production}
\label{sec:assoc}
\begin{figure}[!t]
\begin{center}
\includegraphics[width=0.5\textwidth]{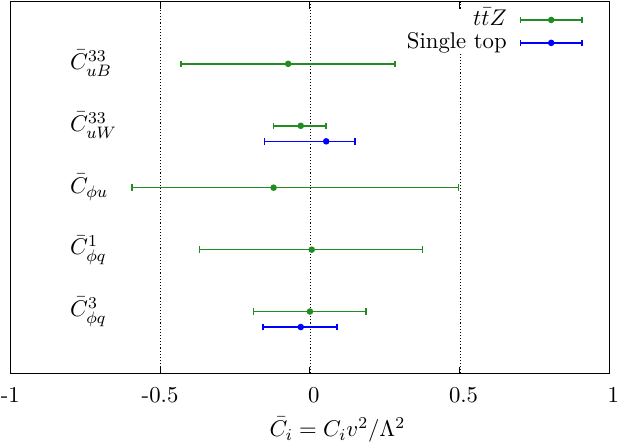}
\caption{Individual 95\% confidence intervals for the operators of \ref{eqn:ttzops} from $t\bar{t}\gamma$ and $t\bar{t}Z$ production (green) and in the two cases where there is overlap, from single top measurements (blue). }
\label{fig:ttzconstraints}
\end{center}
\end{figure}

In addition to top pair and single top production, first measurements have been
reported~\cite{Aad:2015uwa,Aad:2015eua,Khachatryan:2014ewa} of top pair
production in association with a photon and with a $Z$ boson ($t\bar{t}\gamma$
and $t\bar{t}Z$)\footnote{Early measurements of top pair production in
  association with a $W$ has also been reported by ATLAS and CMS, but the
  experimental errors are too large to say anything meaningful about new physics
  therein; the measured cross-sections are still consistent with zero.}. The
cross-section for these processes are considerably smaller, and statistical
uncertainties currently dominate the quoted measurements. Still, they are of
interest because they are sensitive to a new set of operators not previously
accessible, corresponding to enhanced top-gauge couplings which are ubiquitous
in simple $W'$ and $Z$ models, and which allow contact to be made with
electroweak observables. The operator set for $t\bar{t}Z$, for instance,
contains the 6 top pair operators in eq. (\ref{eqn:ttbarops}), plus the
following
\begin{equation}
\begin{split}
\mc{L}_{D6}	& \supset  \frac{C_{uW}}{\Lambda^2} (\bar{q}\sigma^{\mu \nu} \tau^I u)\,\tilde  \varphi \, W_{\mu\nu}^{I} + \frac{C_{uB}}{\Lambda^2} (\bar{q}\sigma^{\mu \nu}  u)\,\tilde \varphi \,B_{\mu\nu}+ \frac{C^{(3)}_{\varphi q}}{\Lambda^2} i(\varphi^\dagger  \overleftrightarrow{D}^I_\mu \varphi )(\bar{q}\gamma^\mu \tau^I q) \\
			&+ \frac{C^{(1)}_{\varphi q}}{\Lambda^2} i(\varphi^\dagger \overleftrightarrow{D}_\mu \varphi )(\bar{q}\gamma^\mu  q) + \frac{C_{\varphi u}}{\Lambda^2}(\varphi^\dagger i \overleftrightarrow{D}_\mu\varphi)(\bar{u}\gamma^\mu u)  \,.
\end{split}
\label{eqn:ttzops}
\end{equation}
There is therefore overlap between the operators contributing to associated production, and those contributing to both top pair and single top. In principle, one should include all observables in a global fit, fitting all coefficients simultaneously. However, the low number of individual $t\bar{t}V$ measurements, coupled with their relatively large uncertainties, means that they do not have much effect on such a fit. Instead, we choose to present individual constraints on the operators from associated production alone, comparing these with top pair and single top in what follows. For the former, we find that the constraints on the operators of eq. (\ref{eqn:ttzops}) obtained from $t\bar{t}\gamma$ and $t\bar{t}Z$ measurements are much weaker than those obtained from top pair production, therefore we do not show them here. The constraints on the new operators of eq.~\eqref{eqn:ttzops} are displayed in Figure~\ref{fig:ttzconstraints}. It is interesting to note that the constraints from associated production measurements are comparable with those from single top production, despite the relative paucity of the former.

\subsection{Decay observables}
\label{sec:decays}
\begin{figure}[t]
\begin{center}
\includegraphics[width=0.5\textwidth]{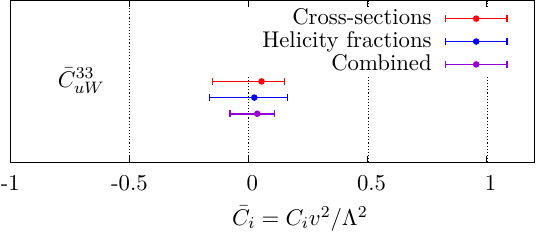}
\caption{95\% bounds on the operator $O_{uW}$ obtained from data on top quark helicity fractions (blue) vs. single top production cross-sections (red), and both sets of measurements combined (purple).  }
\label{fig:helfrac}
\end{center}
\end{figure}

This completes the list of independent dimension-six operators that affect top
quark production cross-sections. However, dimension-six operators may also
contribute (at interference level) to observables relating to top quark decay.
Top quarks decay almost 100\% of the time to a $W$ and $b$ quark. The fraction
of these events which decay to $W$-bosons with a given helicity: left-handed,
right-handed or zero-helicity, can be expressed in terms of helicity fractions,
which for leading order with a finite $b$-quark mass are
\begin{equation}
\begin{split}
F_0 &=\frac{ (1-y^2)^2- x^2(1+y^2)}{(1-y^2)^2+x^2(1-2x^2+y^2)} \\
F_L &= \frac{x^2(1-x^2+y^2)+\sqrt{\lambda}}{(1-y^2)^2+ x^2(1-2x^2 + y^2)} \\
F_R &=  \frac{x^2(1-x^2+y^2) -\sqrt{\lambda}}{(1-y^2)^2+ x^2(1-2x^2 + y^2)}
\end{split}
\end{equation}
where $x=M_W/m_t$, $y=m_b/m_t$ and
$\lambda = 1 + x^4 + y^4 - 2x^2y^2 - 2x^2 - 2y^2 $. As noted
in Ref.~\cite{Zhang:2010dr}, measurements of these fractions can be translated into
bounds on the operator $O_{uW}$. (The operator $O^{(3)}_{\varphi q}$ cannot be accessed in this way, since its only effect is to rescale the
$Wtb$ vertex $V^2_{tb} \to V_{tb} \left(V_{tb}+v^2C^{(3)}_{\varphi q}/\Lambda^2\right)$, therefore it has no effect on
event kinematics.) The desirable feature of
these quantities is that they are relatively stable against higher order
corrections, so the associated scale uncertainties are small. The Standard Model
NNLO estimates for these are: $\{F_0,F_L,F_R \} = \{0.687 \pm 0.005, 0.311
\pm 0.005, 0.0017 \pm 0.0001 \}$~\cite{Czarnecki:2010gb}, i.e. the
uncertainties are at the per mille level. It is interesting to ask whether the
bound obtained on $O_{uW}$ in this way is stronger than that obtained from
cross-section measurements. In Figure~\ref{fig:helfrac} we show the constraints
obtained in each way. Although they are in excellent agreement with each other,
cross-section information gives a slightly stronger bound, mainly due to the
larger amount of data available, but also due to the large experimental
uncertainties on $F_i$. Still, these measurements provide complementary
information on the operator $O_{uW}$, and combining both results in a stronger constraint than either alone, as expected. 

\subsection{Charge asymmetries}
Asymmetries in the production of top quark pairs have received a lot of
attention in recent years, particularly due to an apparent discrepancy between
the Standard Model prediction for the so-called `forward-backward' asymmetry
$\Afb$ in top pair production
\begin{equation}
\Afb = \frac{N(\Delta y > 0)-N(\Delta y < 0)}{N(\Delta y> 0) + N(\Delta y < 0)}
\end{equation}
where $\Delta y = y_t - y_{\bar{t}}$, and a measurement by CDF~\cite{Aaltonen:2011kc}. This discrepancy
was most pronounced in the high invariant mass region, pointing to potential
$\tev$-scale physics at play. However, recent work has cast doubts on its
significance for two reasons: Firstly, an updated analysis with higher
statistics~\cite{Aaltonen:2012it} has slightly lowered the excess. Secondly, a full NNLO QCD
calculation~\cite{Czakon:2014xsa} of $\Afb$ showed that, along with NLO QCD + electroweak
calculations~\cite{Hollik:2011ps,Kuhn:2011ri,Bernreuther:2012sx} the radiative corrections to $\Afb$ are large. The current
measurements are now consistent with the Standard Model within 2$\sigma$. Moreover, the
\dzero experiment reports~\cite{Abazov:2014cca} a high-invariant mass measurement
\textit{lower} than the SM prediction. From a new physics perspective, it is
difficult to accommodate all of this information in a simple,
uncontrived model without tension.

\begin{figure}[!t]
\begin{center}
\includegraphics[width=0.5\textwidth]{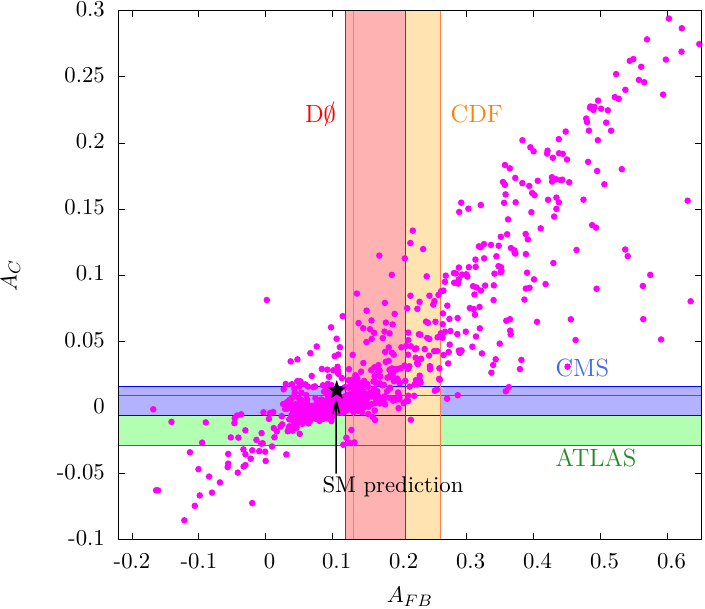}
\caption{Results of a 1000 point parameter space scan over   -10 TeV $^{-2}  < C^{1,2}_{u,d}/\Lambda^2 < $ 10 $TeV ^{-2}$  overlaid with the most up to date measurements of $\Afb$ and \Ac, showing clearly the correlation between them.}
\label{fig:afbvsac}
\end{center}
\end{figure}

Still, in an effective field theory approach, deviations from the Standard Model
prediction of $\Afb$ take a very simple form. A non-zero asymmetry arises from
the difference of four-quark operators:
\begin{equation}
\Afb = (C^1_u-C^2_u+C^1_d-C^2_d)\frac{3\hat{s}\beta}{4g_s^2\Lambda^2(3-\beta^2)},
\end{equation}
where $\beta = \sqrt{1-s/4m_t^2}$ is the velocity of the $t\bar{t}$ system\footnote{Contributions to $A_{FB}$ also arise from the normalisation of $A_{FB}$ and the dimension-six squared term\cite{Bauer:2010iq,AguilarSaavedra:2011vw,Delaunay:2011gv}, which we keep, as discussed in sections 3.3 and 4.}. Combining this inclusive measurement with differential measurements such as $d\Afb/dM_{t\bar{t}}$ allows simultaneous bounds to be extracted on all four of these operators. Therefore it is instructive to compare the bounds obtained on $C^{1,2}_{u,d}$ from charge asymmetries to those obtained from $t\bar{t}$ cross-sections.  Again it is possible to (indirectly) investigate the complementarity between Tevatron
and LHC constraints. Though the charge symmetric initial state of the LHC does
not define a `forward-backward' direction, a related charge asymmetry can be
defined as:
\begin{equation}
A_{C} = \frac{N(\Delta |y| > 0)-N(\Delta |y| < 0)}{N(\Delta |y|> 0) + N(\Delta |y| < 0)}
\end{equation}
making use of the fact that tops tend to be produced at larger rapidities than
antitops. This asymmetry is
diluted with respect to $\Afb$, however. The most up-to-date SM
prediction is $A_C = 0.0123 \pm 0.005$~\cite{Bernreuther:2012sx} for $\sqrt{s} =$ 7 \tev. The experimental status of these measurements is illustrated in Figure~\ref{fig:afbvsac}. The inclusive measurements of
$\Afb$ are consistent with the SM expectation, as are those of
\Ac. The latter, owing to large statistical errors, are also consistent with zero, however, so this result is not particularly conclusive. Since these are different measurements, it is also
possible to modify one without significantly impacting the other. Clearly they
are correlated, as evidenced in Figure~\ref{fig:afbvsac}, where the most up to
date measurements of $\Afb$ and \Ac are shown along with the results
of a 1000 point parameter space scan over the four-quark operators. This
highlights the correlation between the two observables: non-resonant new physics
which causes a large $\Afb$ will also cause a large \Ac, provided it generates a dimension-six operator at low energies.

We have used both inclusive measurements of the charge asymmetries \Ac and $\Afb$, and measurements as a function of the top pair invariant mass $M_{t\bar{t}}$ and rapidity difference $|y_{t\bar{t}}|$. In addition, ATLAS has published measurements of \Ac with a longitudinal `boost' of the $t\bar{t}$ system: $\beta = (|p^z_t+p^z_{\bar{t}})|/(E_t+E_{\bar{t}}) > 0.6 $, which may enhance sensitivity to new physics contributions to \Ac, depending on the model~\cite{AguilarSaavedra:2011cp}.

Since $\Afb = 0$ at leading-order in the SM, it is not possible to define a
$K$-factor in the usual sense. Instead we take higher-order QCD effects into
account by adding the NNLO QCD prediction to the dimension-six terms. In the case of \Ac, we normalise the small (but non-zero) LO QCD piece, to the NLO prediction, which has been calculated with a Monte Carlo and cross-checked with a dedicated NLO calculation~\cite{Bernreuther:2012sx}.

\begin{figure}[t!]
\begin{center}
\includegraphics[width=0.5\textwidth]{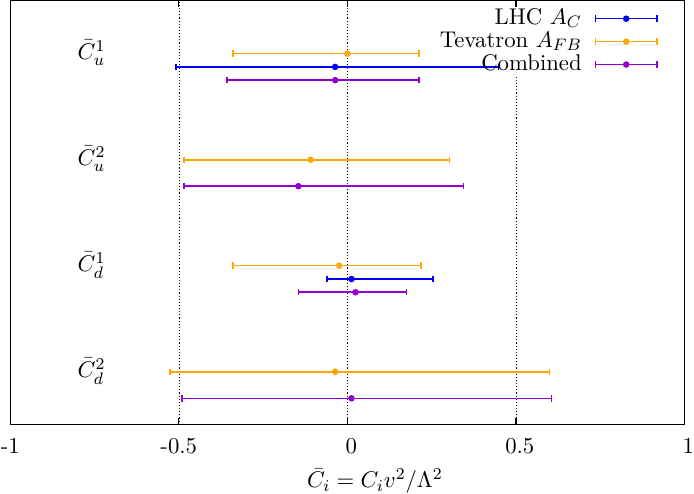}
\caption{Marginalised 95\% confidence intervals on top pair four quark operators from charge asymmetries at the LHC and Tevatron.}
\label{fig:asymms}
\end{center}
\end{figure}

The above asymmetries have been included in the global fit results presented in Figure~\ref{fig:constraints}. However, it is also interesting to see what constraints are obtained on the operators from asymmetry data alone. To this end, the 95\% confidence intervals on the coefficients of the operators $O^{1,2}_{u,d}$ from purely charge
asymmetry data are shown in Figure~\ref{fig:asymms}. Unsurprisingly, the bounds are much weaker than
for cross-section measurements, with the $O^2_i$-type operators unconstrained by LHC data alone.  Despite the small discrepancy between the measured $\Afb$ and its SM value, this does not translate into a non-zero Wilson coefficient; as before, all operators are zero within the 95\% confidence intervals.

At 13~\tev, the asymmetry \Ac will be diluted
even further, due to the increased dominance of the $gg\to t\bar{t}$ channel,
for which $A_C = 0 $. It is therefore possible that charge asymmetry measurements (unlike cross-sections) will not further tighten  the bounds on these operators during LHC Run~II.

\subsection{Contribution of individual datasets}
\begin{figure}[p!]
\begin{center}
\includegraphics[width=0.9\textwidth]{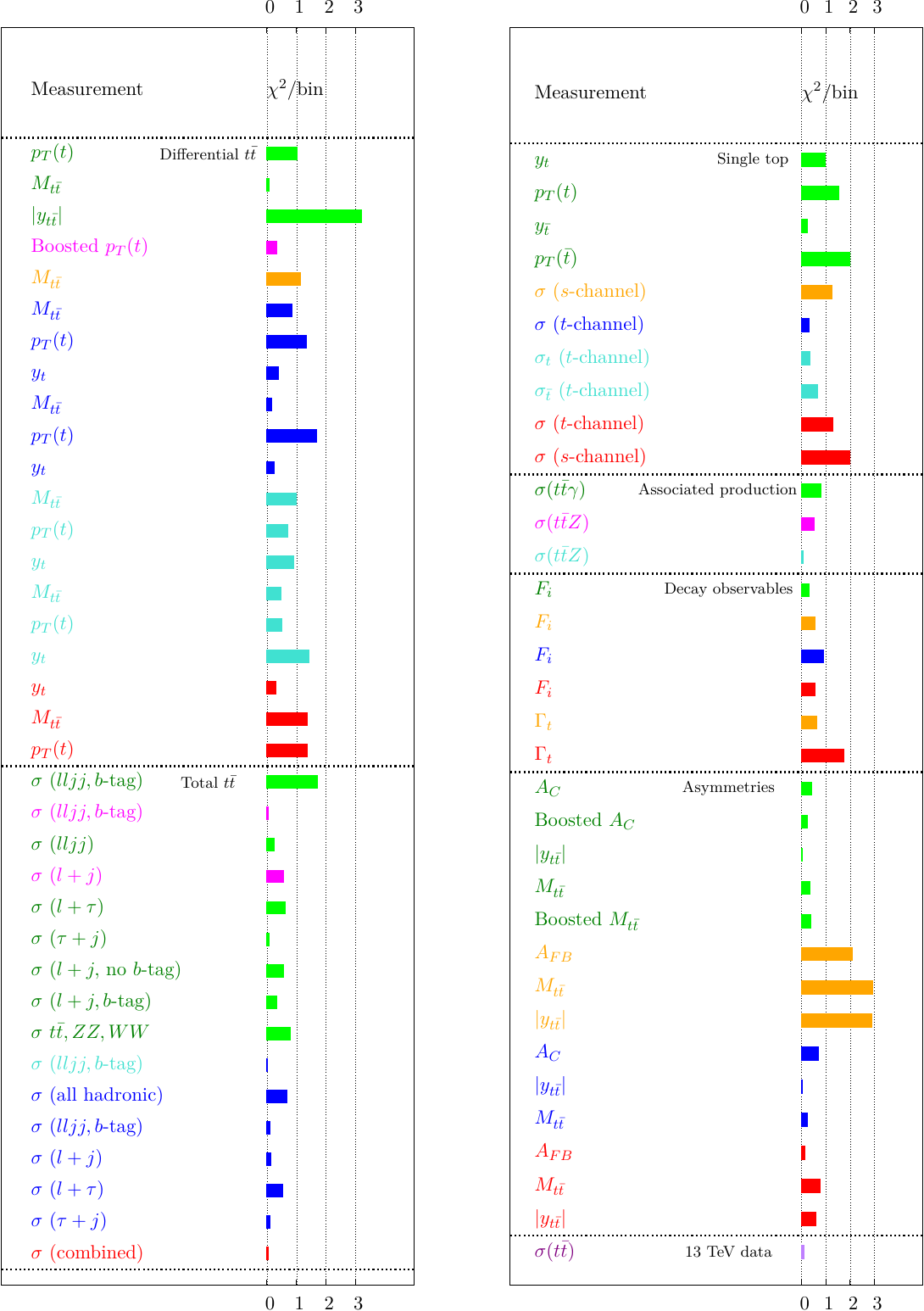}
\caption{$\chi^2$ per bin between measurement and the interpolated best fit point, for measurements considered in this fit. Colours: Green: ATLAS 7 \tev, Magenta: ATLAS 8 \tev, Blue: CMS 7 \tev, Turquoise: CMS 8 \tev, Red: \dzero, Orange: CDF, Purple: CMS 13 \tev.}
\label{fig:pullfactors}
\end{center}
\end{figure}

As well as the constraints presented in Figure~\ref{fig:constraints}, it is also instructive to examine the quality of fit for different datasets. We quantify this by
calculating the $\chi^2$ per bin between the data and the global best fit point,
as shown in Figure~\ref{fig:pullfactors}.

Overall, excellent agreement is seen across the board, with no measurement in obvious tension with any other. The largest single contributors to the $\chi^2$ come from the rapidity distributions in top pair production. It has been known for some time that these are quite poorly modelled with Monte Carlo generators, especially in the boosted regime. It is quite likely that this discrepancy stems from the QCD modelling of the event kinematics, rather than potential new physics. Moreover, in a fit with this many measurements, discrepancies of this magnitude are to be expected on purely statistical grounds.

At the level of total cross-sections, the vanishingly small contributions to the $\chi^2$ stem from two factors: the $\mathcal{O}(10\%)$ measurement uncertainties, which are even larger in hadronic channels, and the large scale uncertainties from the large kinematic range that is integrated over to obtain the total rate. Single top production measurements are also in good agreement with the SM.  The associated production processes $tt\gamma$ and $ttZ$, along with the charge asymmetry measurements from the LHC, have a very small impact on the fit, owing to the large statistical uncertainties on the current measurements. For the former, this situation will improve in Run~II, for the latter the problem will be worse. The forward-backward asymmetry measurements from CDF remain the most discrepant dataset used in the fit.

\section{Constraining UV models}
\label{sec:uvmodels}
As an illustration of the wide-ranging applicability of EFT techniques, we
conclude by matching our effective operator constraints to the low-energy regime
of some specific UV models. These models serve purely illustrative purposes.

\subsection{Axigluon searches}
Considering top pair production, one can imagine the four operators of eq. (\ref{eqn:4fs}) as being generated by integrating out a heavy $s$-channel resonance which interferes with the QCD $q\bar{q}\to t\bar{t}$ amplitude. One particle that could generate such an interference is the so-called axigluon. These originate from models with an extended strong sector with gauge group $SU(3)_{c1}\times SU(3)_{c2}$ which is spontaneously broken to the diagonal subgroup $SU(3)_{c}$ of QCD. In the most minimal scenario, this breaking can be described by a non-linear sigma model
\begin{equation}
\mathcal{L} = -\frac{1}{4}G_{1\mu\nu}G_{1}^{\mu\nu} -\frac{1}{4}G_{2\mu\nu}G_{2}^{\mu\nu} + \frac{f^2}{4}\text{Tr}D_\mu \Sigma D^\mu \Sigma^\dagger \hspace{10pt},\hspace{10pt} \Sigma=\exp\left(\frac{2i\pi^a t^a}{f}\right) \hspace{10pt},\hspace{10pt} a=1,...,8.
\end{equation}
Here $\pi^a$ represent the Goldstone bosons which form the longitudinal degrees of freedom of the colorons, giving them mass, $t^a$ are the Gell-Mann matrices, and $f$ is the symmetry breaking scale. The nonlinear sigma fields transform in the bifundamental representation of $SU(3)_{c1}\times SU(3)_{c2}$:
\begin{equation}
\Sigma \to U_L \Sigma U^\dagger_R \hspace{10pt},\hspace{10pt} U_L = \exp\left(\frac{i\pi^a\alpha_L^a }{f}\right) \hspace{10pt},\hspace{10pt} U_R = \exp\left(\frac{i\pi^a\alpha_R^a }{f}\right)
\end{equation}
The physical fields are obtained by rotating the gauge fields $G_1$ and $G_2$ to the mass eigenstate basis
\begin{equation}
\left(\begin{array}{c} G^a_{1\mu} \\ G^a_{2\mu} \end{array}\right) = \left(\begin{array}{c c} \cos\theta_c &  -\sin\theta_c \\  \sin\theta_c &  \cos\theta_c  \end{array}\right)\left(\begin{array}{c} G^a_\mu \\ C^a_\mu \end{array}\right)
\end{equation}
where the mixing angle $\theta_c$ is defined by
\begin{equation}
\sin\theta_c = \frac{g_{s1}}{\sqrt{g^2_{s1}+g^2_{s2}}}
\end{equation}
The case of an axigluon corresponds to maximal mixing $\theta = \pi/4$, i.e. $g^2_{s1}=g^2_{s2}=g^2_s/2$. Taking the leading-order interference with the SM amplitude for $q\bar{q}\to t\bar{t}$, in the limit $s << M_A^2$, we find that the axigluon induces the dimension-six operators
\begin{equation}
\frac{C^1_u}{\Lambda^2} = \frac{g^2_s}{M_A^2},\hspace{10pt}\hspace{10pt} \frac{C^1_d}{\Lambda^2} = \frac{5g^2_s}{4M_A^2},\hspace{10pt}\hspace{10pt} \frac{C^2_u}{\Lambda^2} =  \frac{C^2_d}{\Lambda^2} = \frac{2 g^2_s}{M_A^2}
\end{equation}
Substituting the marginalised constraints on the 4-quark operators, we find this translates into a lower bound on an axigluon mass. $M_A \gtrsim 1.4 $ \tev at the 95\% confidence level. Since this mass range coincides with the overflow bin of figure~\ref{fig:distributions}, this bound creates some tension with the validity of the EFT approach in the presence of resonances in the $t\bar t$ spectrum (for a general discussion see Ref.~\cite{Englert:2014cva,Brehmer:2015rna,Isidori:2013cga}); at this stage in the LHC programme indirect searches are not sensitive enough to compete with dedicated searches.

\subsection{$W'$ searches}
Turning our attention to single top production, we consider the example of the operator $O^{(3)}_{qq}$ being generated by a heavy charged vector resonance ($W'$) which interferes with the SM amplitude for $s$-channel single top production: $u\bar{d}\to W\to t\bar{b}$. The most general Lagrangian for such a particle (allowing for left and right chiral couplings) is (see e.g. Ref.~\cite{Boos:2006xe}.)
\begin{equation}
\mathcal{L} = \frac{1}{2\sqrt{2}}V_{ij}g_{W'}\bar{q}_i\gamma_\mu(f^R_{ij}(1+\gamma^5)+f^L_{ij}(1-\gamma^5))W^\mu q_j + h.c.
\end{equation}
We take the generic coupling $g_{W'} = g_{SM}$. Since we are considering the interference term only, which must have the same $(V-A)$ structure as the SM, we can set $f^R=0$. Considering the tree-level interference term for between the diagrams for $u\bar{d}\to W', W'\to t\bar{b}$, and taking the limit  $s \ll M_W'^2$ (we also work in the narrow-width approximation $\Gamma_{W}, \Gamma_{W'} \ll M_W, M_{W'} $), we find
\begin{equation}
\label{eq:wprime}
\frac{C^{3,1133}_{qq}}{\Lambda^2} = \frac{g^2}{4M^2_{W'}}
\end{equation}
which, using our global constraint on $O_{t}$, translates into a bound $M_{W'} \gtrsim 1.2 $ \tev.

These bounds are consistent with, but much weaker than, constraints from direct searches for dijet resonances from ATLAS~\cite{Aad:2011fq,Aad:2014aqa} and CMS~\cite{Khachatryan:2015sja}, which report lower bounds of $\{M_A,M_{W'}\} > \{2.72,3.32\} $ \tev and $\{M_A,M_{W'}\} > \{2.2,3.6\} $ \tev respectively. It is unsurprising that these dedicated analyses obtain stronger limits, given the generality of this fit. Again this energy range is resolved in our fit thus in principle invalidating the EFT approach to obtain eq.~\eqref{eq:wprime}. Nonetheless, these bounds provide an interesting comparison of our numerical results, whilst emphasising that for model-specific examples, direct searches for high-mass resonances provide stronger limits than general global fits.

\section{Conclusion}
\label{sec:conclusion}

In this paper, we have performed an up-to-date global fit of top quark effective
field theory to experimental data, including all constrainable operators at
dimension six. For the operators, we use the `Warsaw basis' of
Ref.~\cite{Grzadkowski:2010es}, which has also been widely used in the context
of Higgs and precision electroweak physics. We use data from the Tevatron and
LHC experiments, including LHC Run~II data, up to a centre of mass energy of
13~\tev. Furthermore, we include fully inclusive cross-section measurements, as
well as kinematic distributions involving both the production and decay of the
top quark. Counting each bin independently, the total number of observables
entering our fit is 227, with a total of 13 contributing operators. Constraining
the coefficients of these operators is then a formidable computational task. To
this end we use the parametrisation methods in the \textsc{Professor} framework,
first developed in the context of Monte Carlo generator
tuning~\cite{Buckley:2009bj}, and discussed here in Section~\ref{sec:fit}.

\begin{figure}[!t]
\begin{center}
\includegraphics[width=0.7\textwidth]{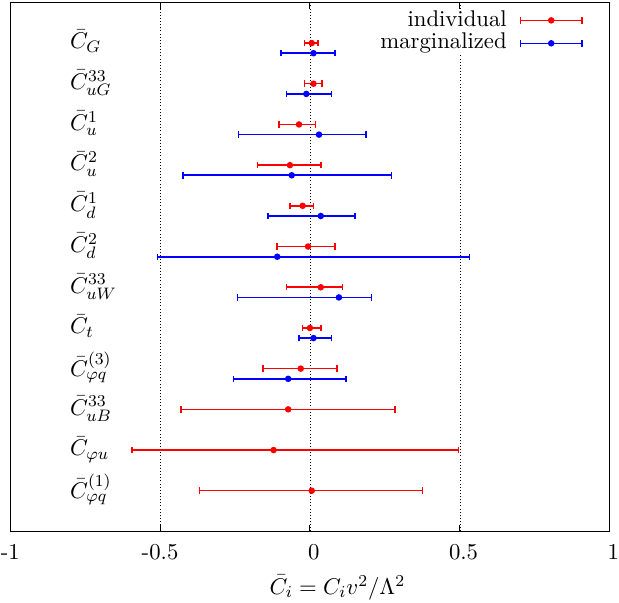}
\caption{95\% confidence intervals for the dimension-six operators that we consider here, with all remaining operators set to zero (red) and marginalised over (blue). In cases where there are constraints on the same operator from different classes of measurement, the strongest limits are shown here. The lack of marginalised constraints for the final three operators is discussed in Section~\ref{sec:assoc}.}
\label{fig:constraints}
\end{center}
\end{figure}

\begin{table}[t!]
\begin{center}
\setlength\extrarowheight{2.5pt}
\begin{tabular}{| c | c | c | } \hline 

 \textbf{Coefficient} & 		 \textbf{Individual constraint} & 		 \textbf{Marginalised constraint} \\ \hline  
$C_{G}v^2/\Lambda^2$ & 		 (---0.018, 0.027) & 		 (---0.097, 0.085) \\ \hline 
$C^{33}_{uG}v^2/\Lambda^2$ & 		 (---0.018, 0.039) & 		 (---0.079, 0.073) \\ \hline 
$C^{1}_{u}v^2/\Lambda^2$ & 		 (---0.103, 0.018) & 		 (---0.236, 0.188) \\ \hline 
$C^{2}_{u}v^2/\Lambda^2$ & 		 (---0.175, 0.036) & 		 (---0.424, 0.272) \\ \hline 
$C^{1}_{d}v^2/\Lambda^2$ & 		 (---0.067, 0.012) & 		 (---0.139, 0.151) \\ \hline 
$C^{2}_{d}v^2/\Lambda^2$ & 		 (---0.109, 0.085) & 		 (---0.508, 0.533) \\ \hline 
$C^{33}_{uW}v^2/\Lambda^2$ & 		 (---0.079, 0.109) & 		 (---0.242, 0.206) \\ \hline 
$C_{t}v^2/\Lambda^2$ & 		 (---0.024, 0.036) & 		 (---0.036, 0.073) \\ \hline 
$C^{3}_{\varphi q}v^2/\Lambda^2$ & 		 (---0.157, 0.091) & 		 (---0.254, 0.121) \\ \hline 
$C^{33}_{uB}v^2/\Lambda^2$ & 		 (---0.430, 0.284) & 		 (---, ---) \\ \hline 
$C_{\varphi u}v^2/\Lambda^2$ & 		 (---0.593, 0.496) & 		 (---, ---) \\ \hline 
$C^{1}_{\varphi q}v^2/\Lambda^2$ & 		 (---0.369, 0.375) & 		 (---, ---) \\ \hline 
\end{tabular}
\caption[95\% confidence intervals on the Wilson coefficients considered in the fit.]{Numerical values of the individual and marginalised 95\% confidence intervals on the operators presented here.} 
\label{table:numbers}
\end{center}
\end{table}

We perform a $\chi^2$ fit of theory to data, including appropriate correlation
matrices where these have been provided by the experiments. We obtain bounds on
the Wilson coefficients of various operators contributing to top quark
production and decay, summarised in Figure~\ref{fig:constraints}, in two cases: (i) when all other coefficients are set to zero; (ii) when all other operators coefficients are marginalised over. The numerical values of these constraints
are also shown in Table~\ref{table:numbers}.

Our stronger constraints are on operators involving the gluon, as expected given the
dominance of gluon fusion in top pair production at the LHC (for which there is
more precise data). Four fermion operators are constrained well in general, with
weaker constraints coming from processes whose experimental uncertainties remain
statistically dominated (e.g. $t\bar{t}V$ production). We have quantified the
interplay between the Tevatron and LHC datasets, as well as that between
different measurement types (e.g. top pair, single top).

Our results currently agree well with the SM only, which is perhaps to be
expected given the lack of reported deviations in previous studies. However, the
fact that this agreement is obtained, in a wide global fit, is itself testament
to the consistency of different top quark measurements, with no obvious tension
between overlapping datasets. There are a number of directions for further
study. Firstly, we can improve the theory description in our fit, to include
higher order QCD corrections in a more rigorous way, as well as moving away from
parton level observables. Secondly, new data from LHC Run~II is continuously
appearing, and can be implemented in our fit as soon as it is available. The era
of performing large global fits to widely different data in the top quark sector
is now upon us, and our work on this area is ongoing.

\section*{Acknowledgments}
We thank Chris Pollard for useful discussions throughout this project. CDW
thanks Andrea Knue for clarifying details of an ATLAS analysis. MR thanks Laure
Berthier and Michael Trott for a helpful discussion, and Alexander Mitov for
correspondence regarding NNLO top pair differential distributions. We are also grateful
 to Ken Mimasu for spotting a typo in Table~\ref{table:numbers}. AB is
supported by a Royal Society University Research Fellowship. DJM, LM, MR and CDW
are supported by the UK Science and Technology Facilities Council (STFC) under
grant ST/L000446/1. JF is supported under STFC grant ST/K001205/1.

\appendix

\bibliographystyle{utphys}
\bibliography{full_ref}

\providecommand{\href}[2]{#2}\begingroup\raggedright\begin{thebibliography}{100}

\bibitem{Appelquist:1974tg}
T.~Appelquist and J.~Carazzone, ``{Infrared Singularities and Massive
  Fields},''
\href{http://dx.doi.org/10.1103/PhysRevD.11.2856}{{\em Phys. Rev.} {\bfseries
  D11} (1975) 2856}.

\bibitem{Wilson:1983dy}
K.~G. Wilson, ``{The renormalization group and critical phenomena},''
\href{http://dx.doi.org/10.1103/RevModPhys.55.583}{{\em Rev. Mod. Phys.}
  {\bfseries 55} (1983) 583--600}.

\bibitem{King:2012tr}
S.~F. King, M.~Muhlleitner, R.~Nevzorov, and K.~Walz, ``{Natural NMSSM Higgs
  Bosons},'' \href{http://dx.doi.org/10.1016/j.nuclphysb.2013.01.020}{{\em
  Nucl. Phys.} {\bfseries B870} (2013) 323--352},
\href{http://arxiv.org/abs/1211.5074}{{\ttfamily arXiv:1211.5074 [hep-ph]}}.

\bibitem{Buchmuller:1985jz}
W.~Buchmuller and D.~Wyler, ``{Effective Lagrangian Analysis of New
  Interactions and Flavor Conservation},''
\href{http://dx.doi.org/10.1016/0550-3213(86)90262-2}{{\em Nucl. Phys.}
  {\bfseries B268} (1986) 621--653}.

\bibitem{Hagiwara:1986vm}
K.~Hagiwara, R.~Peccei, D.~Zeppenfeld, and K.~Hikasa, ``{Probing the Weak Boson
  Sector in $e^+ e^- \to W^+ W^-$},''
\href{http://dx.doi.org/10.1016/0550-3213(87)90685-7}{{\em Nucl.Phys.}
  {\bfseries B282} (1987) 253}.

\bibitem{Burges:1983zg}
C.~J.~C. Burges and H.~J. Schnitzer, ``{Virtual Effects of Excited Quarks as
  Probes of a Possible New Hadronic Mass Scale},''
\href{http://dx.doi.org/10.1016/0550-3213(83)90555-2}{{\em Nucl. Phys.}
  {\bfseries B228} (1983) 464}.

\bibitem{Leung:1984ni}
C.~N. Leung, S.~T. Love, and S.~Rao, ``{Low-Energy Manifestations of a New
  Interaction Scale: Operator Analysis},''
\href{http://dx.doi.org/10.1007/BF01588041}{{\em Z. Phys.} {\bfseries C31}
  (1986) 433}.

\bibitem{Azatov:2012bz}
A.~Azatov, R.~Contino, and J.~Galloway, ``{Model-Independent Bounds on a Light
  Higgs},'' \href{http://dx.doi.org/10.1007/JHEP04(2012)127,
  10.1007/JHEP04(2013)140}{{\em JHEP} {\bfseries 04} (2012) 127},
  \href{http://arxiv.org/abs/1202.3415}{{\ttfamily arXiv:1202.3415 [hep-ph]}}.
[Erratum: JHEP04,140(2013)].

\bibitem{Espinosa:2012im}
J.~R. Espinosa, C.~Grojean, M.~Muhlleitner, and M.~Trott, ``{First Glimpses at
  Higgs' face},'' \href{http://dx.doi.org/10.1007/JHEP12(2012)045}{{\em JHEP}
  {\bfseries 12} (2012) 045},
\href{http://arxiv.org/abs/1207.1717}{{\ttfamily arXiv:1207.1717 [hep-ph]}}.

\bibitem{Plehn:2012iz}
T.~Plehn and M.~Rauch, ``{Higgs Couplings after the Discovery},''
  \href{http://dx.doi.org/10.1209/0295-5075/100/11002}{{\em Europhys. Lett.}
  {\bfseries 100} (2012) 11002},
\href{http://arxiv.org/abs/1207.6108}{{\ttfamily arXiv:1207.6108 [hep-ph]}}.

\bibitem{Carmi:2012in}
D.~Carmi, A.~Falkowski, E.~Kuflik, T.~Volansky, and J.~Zupan, ``{Higgs After
  the Discovery: A Status Report},''
  \href{http://dx.doi.org/10.1007/JHEP10(2012)196}{{\em JHEP} {\bfseries 10}
  (2012) 196},
\href{http://arxiv.org/abs/1207.1718}{{\ttfamily arXiv:1207.1718 [hep-ph]}}.

\bibitem{Peskin:2012we}
M.~E. Peskin, ``{Comparison of LHC and ILC Capabilities for Higgs Boson
  Coupling Measurements},''
\href{http://arxiv.org/abs/1207.2516}{{\ttfamily arXiv:1207.2516 [hep-ph]}}.

\bibitem{Dumont:2013wma}
B.~Dumont, S.~Fichet, and G.~von Gersdorff, ``{A Bayesian view of the Higgs
  sector with higher dimensional operators},''
  \href{http://dx.doi.org/10.1007/JHEP07(2013)065}{{\em JHEP} {\bfseries 07}
  (2013) 065},
\href{http://arxiv.org/abs/1304.3369}{{\ttfamily arXiv:1304.3369 [hep-ph]}}.

\bibitem{Djouadi:2013qya}
A.~Djouadi and G.~Moreau, ``{The couplings of the Higgs boson and its CP
  properties from fits of the signal strengths and their ratios at the 7+8 TeV
  LHC},'' \href{http://dx.doi.org/10.1140/epjc/s10052-013-2512-9}{{\em Eur.
  Phys. J.} {\bfseries C73} no.~9, (2013) 2512},
\href{http://arxiv.org/abs/1303.6591}{{\ttfamily arXiv:1303.6591 [hep-ph]}}.

\bibitem{Lopez-Val:2013yba}
D.~Lopez-Val, T.~Plehn, and M.~Rauch, ``{Measuring extended Higgs sectors as a
  consistent free couplings model},''
  \href{http://dx.doi.org/10.1007/JHEP10(2013)134}{{\em JHEP} {\bfseries 10}
  (2013) 134},
\href{http://arxiv.org/abs/1308.1979}{{\ttfamily arXiv:1308.1979 [hep-ph]}}.

\bibitem{Englert:2014uua}
C.~Englert, A.~Freitas, M.~M. Muehlleitner, T.~Plehn, M.~Rauch, M.~Spira, and
  K.~Walz, ``{Precision Measurements of Higgs Couplings: Implications for New
  Physics Scales},''
  \href{http://dx.doi.org/10.1088/0954-3899/41/11/113001}{{\em J. Phys.}
  {\bfseries G41} (2014) 113001},
\href{http://arxiv.org/abs/1403.7191}{{\ttfamily arXiv:1403.7191 [hep-ph]}}.

\bibitem{Ellis:2014dva}
J.~Ellis, V.~Sanz, and T.~You, ``{Complete Higgs Sector Constraints on
  Dimension-6 Operators},''
  \href{http://dx.doi.org/10.1007/JHEP07(2014)036}{{\em JHEP} {\bfseries 07}
  (2014) 036},
\href{http://arxiv.org/abs/1404.3667}{{\ttfamily arXiv:1404.3667 [hep-ph]}}.

\bibitem{Ellis:2014jta}
J.~Ellis, V.~Sanz, and T.~You, ``{The Effective Standard Model after LHC Run
  I},'' \href{http://dx.doi.org/10.1007/JHEP03(2015)157}{{\em JHEP} {\bfseries
  03} (2015) 157},
\href{http://arxiv.org/abs/1410.7703}{{\ttfamily arXiv:1410.7703 [hep-ph]}}.

\bibitem{Falkowski:2014tna}
A.~Falkowski and F.~Riva, ``{Model-independent precision constraints on
  dimension-6 operators},''
  \href{http://dx.doi.org/10.1007/JHEP02(2015)039}{{\em JHEP} {\bfseries 02}
  (2015) 039},
\href{http://arxiv.org/abs/1411.0669}{{\ttfamily arXiv:1411.0669 [hep-ph]}}.

\bibitem{Corbett:2015ksa}
T.~Corbett, O.~J.~P. Eboli, D.~Goncalves, J.~Gonzalez-Fraile, T.~Plehn, and
  M.~Rauch, ``{The Higgs Legacy of the LHC Run I},''
  \href{http://dx.doi.org/10.1007/JHEP08(2015)156}{{\em JHEP} {\bfseries 08}
  (2015) 156},
\href{http://arxiv.org/abs/1505.05516}{{\ttfamily arXiv:1505.05516 [hep-ph]}}.

\bibitem{Buchalla:2015qju}
G.~Buchalla, O.~Cata, A.~Celis, and C.~Krause, ``{Fitting Higgs Data with
  Nonlinear Effective Theory},''
\href{http://arxiv.org/abs/1511.00988}{{\ttfamily arXiv:1511.00988 [hep-ph]}}.

\bibitem{Aad:2015tna}
{\bfseries ATLAS} Collaboration, G.~Aad {\em et~al.}, ``{Constraints on
  non-Standard Model Higgs boson interactions in an effective field theory
  using differential cross sections measured in the $H \rightarrow
  \gamma\gamma$ decay channel at $\sqrt{s} = 8$ TeV with the ATLAS detector},''
\href{http://arxiv.org/abs/1508.02507}{{\ttfamily arXiv:1508.02507 [hep-ex]}}.

\bibitem{Berthier:2015gja}
L.~Berthier and M.~Trott, ``{Consistent constraints on the Standard Model
  Effective Field Theory},''
\href{http://arxiv.org/abs/1508.05060}{{\ttfamily arXiv:1508.05060 [hep-ph]}}.

\bibitem{Falkowski:2015jaa}
A.~Falkowski, M.~Gonzalez-Alonso, A.~Greljo, and D.~Marzocca, ``{Global
  constraints on anomalous triple gauge couplings in effective field theory
  approach},''
\href{http://arxiv.org/abs/1508.00581}{{\ttfamily arXiv:1508.00581 [hep-ph]}}.

\bibitem{Englert:2015hrx}
C.~Englert, R.~Kogler, H.~Schulz, and M.~Spannowsky, ``{Higgs coupling
  measurements at the LHC},''
\href{http://arxiv.org/abs/1511.05170}{{\ttfamily arXiv:1511.05170 [hep-ph]}}.

\bibitem{AguilarSaavedra:2008zc}
J.~A. Aguilar-Saavedra, ``{A Minimal set of top anomalous couplings},''
  \href{http://dx.doi.org/10.1016/j.nuclphysb.2008.12.012}{{\em Nucl. Phys.}
  {\bfseries B812} (2009) 181--204},
\href{http://arxiv.org/abs/0811.3842}{{\ttfamily arXiv:0811.3842 [hep-ph]}}.

\bibitem{Bernardo:2014vha}
C.~Bernardo, N.~F. Castro, M.~C.~N. Fiolhais, H.~Gonçalves, A.~G.~C. Guerra,
  {\em et~al.}, ``{Studying the $Wtb$ vertex structure using recent LHC
  results},'' \href{http://dx.doi.org/10.1103/PhysRevD.90.113007}{{\em
  Phys.Rev.} {\bfseries D90} no.~11, (2014) 113007},
\href{http://arxiv.org/abs/1408.7063}{{\ttfamily arXiv:1408.7063 [hep-ph]}}.

\bibitem{Grzadkowski:2003tf}
B.~Grzadkowski, Z.~Hioki, K.~Ohkuma, and J.~Wudka, ``{Probing anomalous top
  quark couplings induced by dimension-six operators at photon colliders},''
  \href{http://dx.doi.org/10.1016/j.nuclphysb.2004.04.006}{{\em Nucl. Phys.}
  {\bfseries B689} (2004) 108--126},
\href{http://arxiv.org/abs/hep-ph/0310159}{{\ttfamily arXiv:hep-ph/0310159
  [hep-ph]}}.

\bibitem{Nomura:2009tw}
D.~Nomura, ``{Effects of Top-quark Compositeness on Higgs Boson Production at
  the LHC},'' \href{http://dx.doi.org/10.1007/JHEP02(2010)061}{{\em JHEP}
  {\bfseries 02} (2010) 061},
\href{http://arxiv.org/abs/0911.1941}{{\ttfamily arXiv:0911.1941 [hep-ph]}}.

\bibitem{Hioki:2009hm}
Z.~Hioki and K.~Ohkuma, ``{Search for anomalous top-gluon couplings at LHC
  revisited},'' \href{http://dx.doi.org/10.1140/epjc/s10052-009-1204-y}{{\em
  Eur. Phys. J.} {\bfseries C65} (2010) 127--135},
\href{http://arxiv.org/abs/0910.3049}{{\ttfamily arXiv:0910.3049 [hep-ph]}}.

\bibitem{Hioki:2010zu}
Z.~Hioki and K.~Ohkuma, ``{Addendum to: Search for anomalous top-gluon
  couplings at LHC revisited},''
  \href{http://dx.doi.org/10.1140/epjc/s10052-010-1535-8}{{\em Eur. Phys. J.}
  {\bfseries C71} (2011) 1535},
\href{http://arxiv.org/abs/1011.2655}{{\ttfamily arXiv:1011.2655 [hep-ph]}}.

\bibitem{Hioki:2013hva}
Z.~Hioki and K.~Ohkuma, ``{Latest constraint on nonstandard top-gluon couplings
  at hadron colliders and its future prospect},''
  \href{http://dx.doi.org/10.1103/PhysRevD.88.017503}{{\em Phys. Rev.}
  {\bfseries D88} (2013) 017503},
\href{http://arxiv.org/abs/1306.5387}{{\ttfamily arXiv:1306.5387 [hep-ph]}}.

\bibitem{Aguilar-Saavedra:2014iga}
J.~A. Aguilar-Saavedra, B.~Fuks, and M.~L. Mangano, ``{Pinning down top dipole
  moments with ultra-boosted tops},''
  \href{http://dx.doi.org/10.1103/PhysRevD.91.094021}{{\em Phys. Rev.}
  {\bfseries D91} (2015) 094021},
\href{http://arxiv.org/abs/1412.6654}{{\ttfamily arXiv:1412.6654 [hep-ph]}}.

\bibitem{Chen:2005vr}
C.-R. Chen, F.~Larios, and C.~P. Yuan, ``{General analysis of single top
  production and $W$ helicity in top decay},''
  \href{http://dx.doi.org/10.1016/j.physletb.2005.10.002}{{\em Phys. Lett.}
  {\bfseries B631} (2005) 126--132},
\href{http://arxiv.org/abs/hep-ph/0503040}{{\ttfamily arXiv:hep-ph/0503040
  [hep-ph]}}.

\bibitem{AguilarSaavedra:2008gt}
J.~A. Aguilar-Saavedra, ``{Single top quark production at LHC with anomalous
  Wtb couplings},''
  \href{http://dx.doi.org/10.1016/j.nuclphysb.2008.06.013}{{\em Nucl. Phys.}
  {\bfseries B804} (2008) 160--192},
\href{http://arxiv.org/abs/0803.3810}{{\ttfamily arXiv:0803.3810 [hep-ph]}}.

\bibitem{AguilarSaavedra:2010nx}
J.~A. Aguilar-Saavedra and J.~Bernabeu, ``{W polarisation beyond helicity
  fractions in top quark decays},''
  \href{http://dx.doi.org/10.1016/j.nuclphysb.2010.07.012}{{\em Nucl. Phys.}
  {\bfseries B840} (2010) 349--378},
\href{http://arxiv.org/abs/1005.5382}{{\ttfamily arXiv:1005.5382 [hep-ph]}}.

\bibitem{AguilarSaavedra:2011ct}
J.~A. Aguilar-Saavedra, N.~F. Castro, and A.~Onofre, ``{Constraints on the Wtb
  vertex from early LHC data},''
  \href{http://dx.doi.org/10.1103/PhysRevD.84.019901,
  10.1103/PhysRevD.83.117301}{{\em Phys. Rev.} {\bfseries D83} (2011) 117301},
\href{http://arxiv.org/abs/1105.0117}{{\ttfamily arXiv:1105.0117 [hep-ph]}}.

\bibitem{Fabbrichesi:2014wva}
M.~Fabbrichesi, M.~Pinamonti, and A.~Tonero, ``{Limits on anomalous top quark
  gauge couplings from Tevatron and LHC data},''
  \href{http://dx.doi.org/10.1140/epjc/s10052-014-3193-8}{{\em Eur. Phys. J.}
  {\bfseries C74} no.~12, (2014) 3193},
\href{http://arxiv.org/abs/1406.5393}{{\ttfamily arXiv:1406.5393 [hep-ph]}}.

\bibitem{Fabbrichesi:2013bca}
M.~Fabbrichesi, M.~Pinamonti, and A.~Tonero, ``{Stringent limits on top-quark
  compositeness from $t \bar t$ production at the Tevatron and the LHC},''
  \href{http://dx.doi.org/10.1103/PhysRevD.89.074028}{{\em Phys. Rev.}
  {\bfseries D89} no.~7, (2014) 074028},
\href{http://arxiv.org/abs/1307.5750}{{\ttfamily arXiv:1307.5750 [hep-ph]}}.

\bibitem{Cao:2015doa}
Q.-H. Cao, B.~Yan, J.-H. Yu, and C.~Zhang, ``{A General Analysis of $Wtb$
  anomalous Couplings},''
\href{http://arxiv.org/abs/1504.03785}{{\ttfamily arXiv:1504.03785 [hep-ph]}}.

\bibitem{GonzalezSprinberg:2011kx}
G.~A. Gonzalez-Sprinberg, R.~Martinez, and J.~Vidal, ``{Top quark tensor
  couplings},'' \href{http://dx.doi.org/10.1007/JHEP07(2011)094,
  10.1007/JHEP05(2013)117}{{\em JHEP} {\bfseries 07} (2011) 094},
  \href{http://arxiv.org/abs/1105.5601}{{\ttfamily arXiv:1105.5601 [hep-ph]}}.
[Erratum: JHEP05,117(2013)].

\bibitem{Davidson:2015zza}
S.~Davidson, M.~L. Mangano, S.~Perries, and V.~Sordini, ``{Lepton Flavour
  Violating top decays at the LHC},''
  \href{http://dx.doi.org/10.1140/epjc/s10052-015-3649-5}{{\em Eur. Phys. J.}
  {\bfseries C75} no.~9, (2015) 450},
\href{http://arxiv.org/abs/1507.07163}{{\ttfamily arXiv:1507.07163 [hep-ph]}}.

\bibitem{Jung:2014kxa}
S.~Jung, P.~Ko, Y.~W. Yoon, and C.~Yu, ``{Renormalization group-induced
  phenomena of top pairs from four-quark effective operators},''
  \href{http://dx.doi.org/10.1007/JHEP08(2014)120}{{\em JHEP} {\bfseries 08}
  (2014) 120},
\href{http://arxiv.org/abs/1406.4570}{{\ttfamily arXiv:1406.4570 [hep-ph]}}.

\bibitem{Cao:2007ea}
Q.-H. Cao, J.~Wudka, and C.~P. Yuan, ``{Search for new physics via single top
  production at the LHC},''
  \href{http://dx.doi.org/10.1016/j.physletb.2007.10.057}{{\em Phys. Lett.}
  {\bfseries B658} (2007) 50--56},
\href{http://arxiv.org/abs/0704.2809}{{\ttfamily arXiv:0704.2809 [hep-ph]}}.

\bibitem{Degrande:2010kt}
C.~Degrande, J.-M. Gerard, C.~Grojean, F.~Maltoni, and G.~Servant,
  ``{Non-resonant New Physics in Top Pair Production at Hadron Colliders},''
  \href{http://dx.doi.org/10.1007/JHEP03(2011)125}{{\em JHEP} {\bfseries 03}
  (2011) 125},
\href{http://arxiv.org/abs/1010.6304}{{\ttfamily arXiv:1010.6304 [hep-ph]}}.

\bibitem{Zhang:2010dr}
C.~Zhang and S.~Willenbrock, ``{Effective-Field-Theory Approach to Top-Quark
  Production and Decay},''
  \href{http://dx.doi.org/10.1103/PhysRevD.83.034006}{{\em Phys. Rev.}
  {\bfseries D83} (2011) 034006},
\href{http://arxiv.org/abs/1008.3869}{{\ttfamily arXiv:1008.3869 [hep-ph]}}.

\bibitem{Greiner:2011tt}
N.~Greiner, S.~Willenbrock, and C.~Zhang, ``{Effective Field Theory for
  Nonstandard Top Quark Couplings},''
  \href{http://dx.doi.org/10.1016/j.physletb.2011.09.026}{{\em Phys. Lett.}
  {\bfseries B704} (2011) 218--222},
\href{http://arxiv.org/abs/1104.3122}{{\ttfamily arXiv:1104.3122 [hep-ph]}}.

\bibitem{Degrande:2012wf}
C.~Degrande, N.~Greiner, W.~Kilian, O.~Mattelaer, H.~Mebane, T.~Stelzer,
  S.~Willenbrock, and C.~Zhang, ``{Effective Field Theory: A Modern Approach to
  Anomalous Couplings},''
  \href{http://dx.doi.org/10.1016/j.aop.2013.04.016}{{\em Annals Phys.}
  {\bfseries 335} (2013) 21--32},
\href{http://arxiv.org/abs/1205.4231}{{\ttfamily arXiv:1205.4231 [hep-ph]}}.

\bibitem{deBlas:2015aea}
J.~de~Blas, M.~Chala, and J.~Santiago, ``{Renormalization Group Constraints on
  New Top Interactions from Electroweak Precision Data},''
  \href{http://dx.doi.org/10.1007/JHEP09(2015)189}{{\em JHEP} {\bfseries 09}
  (2015) 189},
\href{http://arxiv.org/abs/1507.00757}{{\ttfamily arXiv:1507.00757 [hep-ph]}}.

\bibitem{Aguilar:2015vsa}
R.~R. Aguilar, A.~O. Bouzas, and F.~Larios, ``{Limits on the anomalous $Wtq$
  couplings},''
\href{http://arxiv.org/abs/1509.06431}{{\ttfamily arXiv:1509.06431 [hep-ph]}}.

\bibitem{Durieux:2014xla}
G.~Durieux, F.~Maltoni, and C.~Zhang, ``{Global approach to top-quark
  flavor-changing interactions},''
  \href{http://dx.doi.org/10.1103/PhysRevD.91.074017}{{\em Phys. Rev.}
  {\bfseries D91} no.~7, (2015) 074017},
\href{http://arxiv.org/abs/1412.7166}{{\ttfamily arXiv:1412.7166 [hep-ph]}}.

\bibitem{Buckley:2015nca}
A.~Buckley, C.~Englert, J.~Ferrando, D.~J. Miller, L.~Moore, M.~Russell, and
  C.~D. White, ``{Global fit of top quark effective theory to data},''
  \href{http://dx.doi.org/10.1103/PhysRevD.92.091501}{{\em Phys. Rev.}
  {\bfseries D92} no.~9, (2015) 091501},
\href{http://arxiv.org/abs/1506.08845}{{\ttfamily arXiv:1506.08845 [hep-ph]}}.

\bibitem{Buckley:2009bj}
A.~Buckley, H.~Hoeth, H.~Lacker, H.~Schulz, and J.~E. von Seggern,
  ``{Systematic event generator tuning for the LHC},''
  \href{http://dx.doi.org/10.1140/epjc/s10052-009-1196-7}{{\em Eur.Phys.J.}
  {\bfseries C65} (2010) 331--357},
\href{http://arxiv.org/abs/0907.2973}{{\ttfamily arXiv:0907.2973 [hep-ph]}}.

\bibitem{Grzadkowski:2010es}
B.~Grzadkowski, M.~Iskrzynski, M.~Misiak, and J.~Rosiek, ``{Dimension-Six Terms
  in the Standard Model Lagrangian},''
  \href{http://dx.doi.org/10.1007/JHEP10(2010)085}{{\em JHEP} {\bfseries 10}
  (2010) 085},
\href{http://arxiv.org/abs/1008.4884}{{\ttfamily arXiv:1008.4884 [hep-ph]}}.

\bibitem{Gupta:2014rxa}
R.~S. Gupta, A.~Pomarol, and F.~Riva, ``{BSM Primary Effects},''
  \href{http://dx.doi.org/10.1103/PhysRevD.91.035001}{{\em Phys. Rev.}
  {\bfseries D91} no.~3, (2015) 035001},
\href{http://arxiv.org/abs/1405.0181}{{\ttfamily arXiv:1405.0181 [hep-ph]}}.

\bibitem{Giudice:2007fh}
G.~F. Giudice, C.~Grojean, A.~Pomarol, and R.~Rattazzi, ``{The
  Strongly-Interacting Light Higgs},''
  \href{http://dx.doi.org/10.1088/1126-6708/2007/06/045}{{\em JHEP} {\bfseries
  06} (2007) 045},
\href{http://arxiv.org/abs/hep-ph/0703164}{{\ttfamily arXiv:hep-ph/0703164
  [hep-ph]}}.

\bibitem{Contino:2013kra}
R.~Contino, M.~Ghezzi, C.~Grojean, M.~Muhlleitner, and M.~Spira, ``{Effective
  Lagrangian for a light Higgs-like scalar},''
  \href{http://dx.doi.org/10.1007/JHEP07(2013)035}{{\em JHEP} {\bfseries 07}
  (2013) 035},
\href{http://arxiv.org/abs/1303.3876}{{\ttfamily arXiv:1303.3876 [hep-ph]}}.

\bibitem{Masso:2014xra}
E.~Masso, ``{An Effective Guide to Beyond the Standard Model Physics},''
  \href{http://dx.doi.org/10.1007/JHEP10(2014)128}{{\em JHEP} {\bfseries 10}
  (2014) 128},
\href{http://arxiv.org/abs/1406.6376}{{\ttfamily arXiv:1406.6376 [hep-ph]}}.

\bibitem{Pomarol:2014dya}
A.~Pomarol, ``{Higgs Physics},'' in {\em {2014 European School of High-Energy
  Physics (ESHEP 2014) Garderen, The Netherlands, June 18-July 1, 2014}}.
\newblock 2014.
\newblock
\href{http://arxiv.org/abs/1412.4410}{{\ttfamily arXiv:1412.4410 [hep-ph]}}.
\newblock

\bibitem{Falkowski:2015wza}
A.~Falkowski, B.~Fuks, K.~Mawatari, K.~Mimasu, F.~Riva, and V.~sanz,
  ``{Rosetta: an operator basis translator for Standard Model effective field
  theory},''
\href{http://arxiv.org/abs/1508.05895}{{\ttfamily arXiv:1508.05895 [hep-ph]}}.

\bibitem{Chatrchyan:2013wua}
{\bfseries CMS} Collaboration, S.~Chatrchyan {\em et~al.}, ``{Measurements of
  $t\bar{t}$ spin correlations and top-quark polarization using dilepton final
  states in $pp$ collisions at $\sqrt{s}$ = 7 TeV},''
  \href{http://dx.doi.org/10.1103/PhysRevLett.112.182001}{{\em Phys. Rev.
  Lett.} {\bfseries 112} no.~18, (2014) 182001},
\href{http://arxiv.org/abs/1311.3924}{{\ttfamily arXiv:1311.3924 [hep-ex]}}.

\bibitem{Aad:2014pwa}
{\bfseries ATLAS} Collaboration, G.~Aad {\em et~al.}, ``{Measurements of spin
  correlation in top-antitop quark events from proton-proton collisions at
  $\sqrt{s}=7$ TeV using the ATLAS detector},''
  \href{http://dx.doi.org/10.1103/PhysRevD.90.112016}{{\em Phys. Rev.}
  {\bfseries D90} no.~11, (2014) 112016},
\href{http://arxiv.org/abs/1407.4314}{{\ttfamily arXiv:1407.4314 [hep-ex]}}.

\bibitem{Aad:2014kva}
{\bfseries ATLAS} Collaboration, G.~Aad {\em et~al.}, ``{Measurement of the
  $t\overline{t}$ production cross-section using $e\mu $ events with $b$
  -tagged jets in $pp$ collisions at $\sqrt{s}=7$ and 8 TeV with the ATLAS
  detector},'' \href{http://dx.doi.org/10.1140/epjc/s10052-014-3109-7}{{\em
  Eur. Phys. J.} {\bfseries C74} no.~10, (2014) 3109},
\href{http://arxiv.org/abs/1406.5375}{{\ttfamily arXiv:1406.5375 [hep-ex]}}.

\bibitem{ATLAS:2012aa}
{\bfseries ATLAS} Collaboration, G.~Aad {\em et~al.}, ``{Measurement of the
  cross section for top-quark pair production in $pp$ collisions at
  $\sqrt{s}=7$ TeV with the ATLAS detector using final states with two high-pt
  leptons},'' \href{http://dx.doi.org/10.1007/JHEP05(2012)059}{{\em JHEP}
  {\bfseries 05} (2012) 059},
\href{http://arxiv.org/abs/1202.4892}{{\ttfamily arXiv:1202.4892 [hep-ex]}}.

\bibitem{Aad:2012mza}
{\bfseries ATLAS} Collaboration, G.~Aad {\em et~al.}, ``{Measurement of the top
  quark pair cross section with ATLAS in $pp$ collisions at $\sqrt{s} =$ 7 TeV
  using final states with an electron or a muon and a hadronically decaying
  $\tau$ lepton},''
  \href{http://dx.doi.org/10.1016/j.physletb.2012.09.032}{{\em Phys. Lett.}
  {\bfseries B717} (2012) 89--108},
\href{http://arxiv.org/abs/1205.2067}{{\ttfamily arXiv:1205.2067 [hep-ex]}}.

\bibitem{Aad:2012qf}
{\bfseries ATLAS} Collaboration, G.~Aad {\em et~al.}, ``{Measurement of the top
  quark pair production cross-section with ATLAS in the single lepton
  channel},'' \href{http://dx.doi.org/10.1016/j.physletb.2012.03.083}{{\em
  Phys. Lett.} {\bfseries B711} (2012) 244--263},
\href{http://arxiv.org/abs/1201.1889}{{\ttfamily arXiv:1201.1889 [hep-ex]}}.

\bibitem{Aad:2012vip}
{\bfseries ATLAS} Collaboration, G.~Aad {\em et~al.}, ``{Measurement of the
  ttbar production cross section in the tau+jets channel using the ATLAS
  detector},'' \href{http://dx.doi.org/10.1140/epjc/s10052-013-2328-7}{{\em
  Eur. Phys. J.} {\bfseries C73} no.~3, (2013) 2328},
\href{http://arxiv.org/abs/1211.7205}{{\ttfamily arXiv:1211.7205 [hep-ex]}}.

\bibitem{Aad:2014jra}
{\bfseries ATLAS} Collaboration, G.~Aad {\em et~al.}, ``{Simultaneous
  measurements of the $t\bar{t}$, $W^+W^-$, and
  $Z/\gamma^{*}\rightarrow\tau\tau$ production cross-sections in $pp$
  collisions at $\sqrt{s} = 7$ TeV with the ATLAS detector},''
  \href{http://dx.doi.org/10.1103/PhysRevD.91.052005}{{\em Phys. Rev.}
  {\bfseries D91} no.~5, (2015) 052005},
\href{http://arxiv.org/abs/1407.0573}{{\ttfamily arXiv:1407.0573 [hep-ex]}}.

\bibitem{Aad:2015pga}
{\bfseries ATLAS} Collaboration, G.~Aad {\em et~al.}, ``{Measurement of the top
  pair production cross section in 8 TeV proton-proton collisions using
  kinematic information in the lepton+jets final state with ATLAS},''
  \href{http://dx.doi.org/10.1103/PhysRevD.91.112013}{{\em Phys. Rev.}
  {\bfseries D91} no.~11, (2015) 112013},
\href{http://arxiv.org/abs/1504.04251}{{\ttfamily arXiv:1504.04251 [hep-ex]}}.

\bibitem{Chatrchyan:2013ual}
{\bfseries CMS} Collaboration, S.~Chatrchyan {\em et~al.}, ``{Measurement of
  the $t\bar{t}$ production cross section in the all-jet final state in pp
  collisions at $\sqrt{s}$ = 7 TeV},''
  \href{http://dx.doi.org/10.1007/JHEP05(2013)065}{{\em JHEP} {\bfseries 05}
  (2013) 065},
\href{http://arxiv.org/abs/1302.0508}{{\ttfamily arXiv:1302.0508 [hep-ex]}}.

\bibitem{Chatrchyan:2012bra}
{\bfseries CMS} Collaboration, S.~Chatrchyan {\em et~al.}, ``{Measurement of
  the $t\bar{t}$ production cross section in the dilepton channel in $pp$
  collisions at $\sqrt{s}=7$ TeV},''
  \href{http://dx.doi.org/10.1007/JHEP11(2012)067}{{\em JHEP} {\bfseries 11}
  (2012) 067},
\href{http://arxiv.org/abs/1208.2671}{{\ttfamily arXiv:1208.2671 [hep-ex]}}.

\bibitem{Chatrchyan:2012ria}
{\bfseries CMS} Collaboration, S.~Chatrchyan {\em et~al.}, ``{Measurement of
  the $t\bar{t}$ production cross section in $pp$ collisions at $\sqrt{s}=7$
  TeV with lepton + jets final states},''
  \href{http://dx.doi.org/10.1016/j.physletb.2013.02.021}{{\em Phys. Lett.}
  {\bfseries B720} (2013) 83--104},
\href{http://arxiv.org/abs/1212.6682}{{\ttfamily arXiv:1212.6682}}.

\bibitem{Chatrchyan:2012vs}
{\bfseries CMS} Collaboration, S.~Chatrchyan {\em et~al.}, ``{Measurement of
  the top quark pair production cross section in $pp$ collisions at $\sqrt{s} =
  7$ TeV in dilepton final states containing a $\tau$},''
  \href{http://dx.doi.org/10.1103/PhysRevD.85.112007}{{\em Phys. Rev.}
  {\bfseries D85} (2012) 112007},
\href{http://arxiv.org/abs/1203.6810}{{\ttfamily arXiv:1203.6810 [hep-ex]}}.

\bibitem{Chatrchyan:2013kff}
{\bfseries CMS} Collaboration, S.~Chatrchyan {\em et~al.}, ``{Measurement of
  the top-antitop production cross section in the tau+jets channel in pp
  collisions at sqrt(s) = 7 TeV},''
  \href{http://dx.doi.org/10.1140/epjc/s10052-013-2386-x}{{\em Eur. Phys. J.}
  {\bfseries C73} no.~4, (2013) 2386},
\href{http://arxiv.org/abs/1301.5755}{{\ttfamily arXiv:1301.5755 [hep-ex]}}.

\bibitem{Chatrchyan:2013faa}
{\bfseries CMS} Collaboration, S.~Chatrchyan {\em et~al.}, ``{Measurement of
  the $t \bar{t}$ production cross section in the dilepton channel in pp
  collisions at $\sqrt{s}$ = 8 TeV},''
  \href{http://dx.doi.org/10.1007/JHEP02(2014)024,
  10.1007/JHEP02(2014)102}{{\em JHEP} {\bfseries 02} (2014) 024},
  \href{http://arxiv.org/abs/1312.7582}{{\ttfamily arXiv:1312.7582 [hep-ex]}}.
[Erratum: JHEP02,102(2014)].

\bibitem{Khachatryan:2015uqb}
{\bfseries CMS} Collaboration, V.~Khachatryan {\em et~al.}, ``{Measurement of
  the top quark pair production cross section in proton-proton collisions at
  $\sqrt{s}=13$ TeV},''
\href{http://arxiv.org/abs/1510.05302}{{\ttfamily arXiv:1510.05302 [hep-ex]}}.

\bibitem{Aaltonen:2013wca}
{\bfseries CDF, D0} Collaboration, T.~A. Aaltonen {\em et~al.}, ``{Combination
  of measurements of the top-quark pair production cross section from the
  Tevatron Collider},''
  \href{http://dx.doi.org/10.1103/PhysRevD.89.072001}{{\em Phys. Rev.}
  {\bfseries D89} no.~7, (2014) 072001},
\href{http://arxiv.org/abs/1309.7570}{{\ttfamily arXiv:1309.7570 [hep-ex]}}.

\bibitem{Aad:2014fwa}
{\bfseries ATLAS} Collaboration, G.~Aad {\em et~al.}, ``{Comprehensive
  measurements of $t$-channel single top-quark production cross sections at
  $\sqrt{s} = 7$ TeV with the ATLAS detector},''
  \href{http://dx.doi.org/10.1103/PhysRevD.90.112006}{{\em Phys. Rev.}
  {\bfseries D90} no.~11, (2014) 112006},
\href{http://arxiv.org/abs/1406.7844}{{\ttfamily arXiv:1406.7844 [hep-ex]}}.

\bibitem{Aaltonen:2014qja}
{\bfseries CDF} Collaboration, T.~A. Aaltonen {\em et~al.}, ``{Evidence for
  $s$-channel Single-Top-Quark Production in Events with one Charged Lepton and
  two Jets at CDF},''
  \href{http://dx.doi.org/10.1103/PhysRevLett.112.231804}{{\em Phys. Rev.
  Lett.} {\bfseries 112} (2014) 231804},
\href{http://arxiv.org/abs/1402.0484}{{\ttfamily arXiv:1402.0484 [hep-ex]}}.

\bibitem{Khachatryan:2014iya}
{\bfseries CMS} Collaboration, V.~Khachatryan {\em et~al.}, ``{Measurement of
  the t-channel single-top-quark production cross section and of the $\mid
  V_{tb} \mid$ CKM matrix element in pp collisions at $\sqrt{s}$= 8 TeV},''
  \href{http://dx.doi.org/10.1007/JHEP06(2014)090}{{\em JHEP} {\bfseries 06}
  (2014) 090},
\href{http://arxiv.org/abs/1403.7366}{{\ttfamily arXiv:1403.7366 [hep-ex]}}.

\bibitem{Abazov:2009pa}
{\bfseries D0} Collaboration, V.~M. Abazov {\em et~al.}, ``{Measurement of the
  t-channel single top quark production cross section},''
  \href{http://dx.doi.org/10.1016/j.physletb.2009.11.038}{{\em Phys. Lett.}
  {\bfseries B682} (2010) 363--369},
\href{http://arxiv.org/abs/0907.4259}{{\ttfamily arXiv:0907.4259 [hep-ex]}}.

\bibitem{Abazov:2011rz}
{\bfseries D0} Collaboration, V.~M. Abazov {\em et~al.}, ``{Model-independent
  measurement of $t$-channel single top quark production in $p\bar{p}$
  collisions at $\sqrt{s}=1.96$ TeV},''
  \href{http://dx.doi.org/10.1016/j.physletb.2011.10.035}{{\em Phys. Lett.}
  {\bfseries B705} (2011) 313--319},
\href{http://arxiv.org/abs/1105.2788}{{\ttfamily arXiv:1105.2788 [hep-ex]}}.

\bibitem{Aad:2014zka}
{\bfseries ATLAS} Collaboration, G.~Aad {\em et~al.}, ``{Measurements of
  normalized differential cross sections for $t\bar{t}$ production in pp
  collisions at $\sqrt{s}=7$ TeV using the ATLAS detector},''
  \href{http://dx.doi.org/10.1103/PhysRevD.90.072004}{{\em Phys. Rev.}
  {\bfseries D90} no.~7, (2014) 072004},
\href{http://arxiv.org/abs/1407.0371}{{\ttfamily arXiv:1407.0371 [hep-ex]}}.

\bibitem{Aaltonen:2009iz}
{\bfseries CDF} Collaboration, T.~Aaltonen {\em et~al.}, ``{First Measurement
  of the t anti-t Differential Cross Section d sigma/dM(t anti-t) in p anti-p
  Collisions at s**(1/2)=1.96-TeV},''
  \href{http://dx.doi.org/10.1103/PhysRevLett.102.222003}{{\em Phys. Rev.
  Lett.} {\bfseries 102} (2009) 222003},
\href{http://arxiv.org/abs/0903.2850}{{\ttfamily arXiv:0903.2850 [hep-ex]}}.

\bibitem{Chatrchyan:2012saa}
{\bfseries CMS} Collaboration, S.~Chatrchyan {\em et~al.}, ``{Measurement of
  differential top-quark pair production cross sections in $pp$ colisions at
  $\sqrt{s}=7$ TeV},''
  \href{http://dx.doi.org/10.1140/epjc/s10052-013-2339-4}{{\em Eur. Phys. J.}
  {\bfseries C73} no.~3, (2013) 2339},
\href{http://arxiv.org/abs/1211.2220}{{\ttfamily arXiv:1211.2220 [hep-ex]}}.

\bibitem{Khachatryan:2015oqa}
{\bfseries CMS} Collaboration, V.~Khachatryan {\em et~al.}, ``{Measurement of
  the Differential Cross Section for Top Quark Pair Production in pp Collisions
  at $\sqrt{s}$ = 8 TeV},''
\href{http://arxiv.org/abs/1505.04480}{{\ttfamily arXiv:1505.04480 [hep-ex]}}.

\bibitem{Abazov:2014vga}
{\bfseries D0} Collaboration, V.~M. Abazov {\em et~al.}, ``{Measurement of
  differential $t\bar{t}$ production cross sections in $p\bar{p}$
  collisions},'' \href{http://dx.doi.org/10.1103/PhysRevD.90.092006}{{\em Phys.
  Rev.} {\bfseries D90} no.~9, (2014) 092006},
\href{http://arxiv.org/abs/1401.5785}{{\ttfamily arXiv:1401.5785 [hep-ex]}}.

\bibitem{Aad:2013cea}
{\bfseries ATLAS} Collaboration, G.~Aad {\em et~al.}, ``{Measurement of the top
  quark pair production charge asymmetry in proton-proton collisions at
  $\sqrt{s}$ = 7 TeV using the ATLAS detector},''
  \href{http://dx.doi.org/10.1007/JHEP02(2014)107}{{\em JHEP} {\bfseries 02}
  (2014) 107},
\href{http://arxiv.org/abs/1311.6724}{{\ttfamily arXiv:1311.6724 [hep-ex]}}.

\bibitem{Chatrchyan:2014yta}
{\bfseries CMS} Collaboration, S.~Chatrchyan {\em et~al.}, ``{Measurements of
  the $t\bar{t}$ charge asymmetry using the dilepton decay channel in pp
  collisions at $\sqrt{s} =$ 7 TeV},''
  \href{http://dx.doi.org/10.1007/JHEP04(2014)191}{{\em JHEP} {\bfseries 04}
  (2014) 191},
\href{http://arxiv.org/abs/1402.3803}{{\ttfamily arXiv:1402.3803 [hep-ex]}}.

\bibitem{Aaltonen:2012it}
{\bfseries CDF} Collaboration, T.~Aaltonen {\em et~al.}, ``{Measurement of the
  top quark forward-backward production asymmetry and its dependence on event
  kinematic properties},''
  \href{http://dx.doi.org/10.1103/PhysRevD.87.092002}{{\em Phys. Rev.}
  {\bfseries D87} no.~9, (2013) 092002},
\href{http://arxiv.org/abs/1211.1003}{{\ttfamily arXiv:1211.1003 [hep-ex]}}.

\bibitem{Abazov:2014cca}
{\bfseries D0} Collaboration, V.~M. Abazov {\em et~al.}, ``{Measurement of the
  forward-backward asymmetry in top quark-antiquark production in ppbar
  collisions using the lepton+jets channel},''
  \href{http://dx.doi.org/10.1103/PhysRevD.90.072011}{{\em Phys. Rev.}
  {\bfseries D90} (2014) 072011},
\href{http://arxiv.org/abs/1405.0421}{{\ttfamily arXiv:1405.0421 [hep-ex]}}.

\bibitem{Aaltonen:2013kna}
{\bfseries CDF} Collaboration, T.~A. Aaltonen {\em et~al.}, ``{Direct
  Measurement of the Total Decay Width of the Top Quark},''
  \href{http://dx.doi.org/10.1103/PhysRevLett.111.202001}{{\em Phys. Rev.
  Lett.} {\bfseries 111} no.~20, (2013) 202001},
\href{http://arxiv.org/abs/1308.4050}{{\ttfamily arXiv:1308.4050 [hep-ex]}}.

\bibitem{Abazov:2012vd}
{\bfseries D0} Collaboration, V.~M. Abazov {\em et~al.}, ``{An Improved
  determination of the width of the top quark},''
  \href{http://dx.doi.org/10.1103/PhysRevD.85.091104}{{\em Phys. Rev.}
  {\bfseries D85} (2012) 091104},
\href{http://arxiv.org/abs/1201.4156}{{\ttfamily arXiv:1201.4156 [hep-ex]}}.

\bibitem{Aad:2012ky}
{\bfseries ATLAS} Collaboration, G.~Aad {\em et~al.}, ``{Measurement of the W
  boson polarization in top quark decays with the ATLAS detector},''
  \href{http://dx.doi.org/10.1007/JHEP06(2012)088}{{\em JHEP} {\bfseries 06}
  (2012) 088},
\href{http://arxiv.org/abs/1205.2484}{{\ttfamily arXiv:1205.2484 [hep-ex]}}.

\bibitem{Aaltonen:2012lua}
{\bfseries CDF} Collaboration, T.~Aaltonen {\em et~al.}, ``{Measurement of
  $W$-Boson Polarization in Top-quark Decay using the Full CDF Run II Data
  Set},'' \href{http://dx.doi.org/10.1103/PhysRevD.87.031104}{{\em Phys. Rev.}
  {\bfseries D87} no.~3, (2013) 031104},
\href{http://arxiv.org/abs/1211.4523}{{\ttfamily arXiv:1211.4523 [hep-ex]}}.

\bibitem{Chatrchyan:2013jna}
{\bfseries CMS} Collaboration, S.~Chatrchyan {\em et~al.}, ``{Measurement of
  the W-boson helicity in top-quark decays from $t\bar{t}$ production in
  lepton+jets events in pp collisions at $\sqrt{s} =$ 7 TeV},''
  \href{http://dx.doi.org/10.1007/JHEP10(2013)167}{{\em JHEP} {\bfseries 10}
  (2013) 167},
\href{http://arxiv.org/abs/1308.3879}{{\ttfamily arXiv:1308.3879 [hep-ex]}}.

\bibitem{Abazov:2010jn}
{\bfseries D0} Collaboration, V.~M. Abazov {\em et~al.}, ``{Measurement of the
  W boson helicity in top quark decays using 5.4 fb$^{-1}$ of $p\bar{p}$
  collision data},'' \href{http://dx.doi.org/10.1103/PhysRevD.83.032009}{{\em
  Phys. Rev.} {\bfseries D83} (2011) 032009},
\href{http://arxiv.org/abs/1011.6549}{{\ttfamily arXiv:1011.6549 [hep-ex]}}.

\bibitem{Aad:2015uwa}
{\bfseries ATLAS} Collaboration, G.~Aad {\em et~al.}, ``{Observation of
  top-quark pair production in association with a photon and measurement of the
  $t\bar{t}\gamma$ production cross section in pp collisions at $\sqrt{s}=7$
  TeV using the ATLAS detector},''
  \href{http://dx.doi.org/10.1103/PhysRevD.91.072007}{{\em Phys. Rev.}
  {\bfseries D91} no.~7, (2015) 072007},
\href{http://arxiv.org/abs/1502.00586}{{\ttfamily arXiv:1502.00586 [hep-ex]}}.

\bibitem{Aad:2015eua}
{\bfseries ATLAS} Collaboration, G.~Aad {\em et~al.}, ``{Measurement of the
  $t\bar{t}W$ and $t\bar{t}Z$ production cross sections in $pp$ collisions at
  $\sqrt{s} = 8 \mathrm{\ Te\kern -0.1em V}$ with the ATLAS detector},''
\href{http://arxiv.org/abs/1509.05276}{{\ttfamily arXiv:1509.05276 [hep-ex]}}.

\bibitem{Khachatryan:2014ewa}
{\bfseries CMS} Collaboration, V.~Khachatryan {\em et~al.}, ``{Measurement of
  top quark-antiquark pair production in association with a W or Z boson in pp
  collisions at $\sqrt{s} = 8$ $\,\text {TeV}$},''
  \href{http://dx.doi.org/10.1140/epjc/s10052-014-3060-7}{{\em Eur. Phys. J.}
  {\bfseries C74} no.~9, (2014) 3060},
\href{http://arxiv.org/abs/1406.7830}{{\ttfamily arXiv:1406.7830 [hep-ex]}}.

\bibitem{Butterworth:2015oua}
J.~Butterworth {\em et~al.}, ``{PDF4LHC recommendations for LHC Run II},''
\href{http://arxiv.org/abs/1510.03865}{{\ttfamily arXiv:1510.03865 [hep-ph]}}.

\bibitem{Nadolsky:2008zw}
P.~M. Nadolsky, H.-L. Lai, Q.-H. Cao, J.~Huston, J.~Pumplin, {\em et~al.},
  ``{Implications of CTEQ global analysis for collider observables},''
  \href{http://dx.doi.org/10.1103/PhysRevD.78.013004}{{\em Phys.Rev.}
  {\bfseries D78} (2008) 013004},
\href{http://arxiv.org/abs/0802.0007}{{\ttfamily arXiv:0802.0007 [hep-ph]}}.

\bibitem{Martin:2009iq}
A.~Martin, W.~Stirling, R.~Thorne, and G.~Watt, ``{Parton distributions for the
  LHC},'' \href{http://dx.doi.org/10.1140/epjc/s10052-009-1072-5}{{\em
  Eur.Phys.J.} {\bfseries C63} (2009) 189--285},
\href{http://arxiv.org/abs/0901.0002}{{\ttfamily arXiv:0901.0002 [hep-ph]}}.

\bibitem{Ball:2010de}
R.~D. Ball, L.~Del~Debbio, S.~Forte, A.~Guffanti, J.~I. Latorre, {\em et~al.},
  ``{A first unbiased global NLO determination of parton distributions and
  their uncertainties},''
  \href{http://dx.doi.org/10.1016/j.nuclphysb.2010.05.008}{{\em Nucl.Phys.}
  {\bfseries B838} (2010) 136--206},
\href{http://arxiv.org/abs/1002.4407}{{\ttfamily arXiv:1002.4407 [hep-ph]}}.

\bibitem{Passarino:2012cb}
G.~Passarino, ``{NLO Inspired Effective Lagrangians for Higgs Physics},''
  \href{http://dx.doi.org/10.1016/j.nuclphysb.2012.11.018}{{\em Nucl. Phys.}
  (2013) 416--458},
\href{http://arxiv.org/abs/1209.5538}{{\ttfamily arXiv:1209.5538 [hep-ph]}}.

\bibitem{Mebane:2013zga}
H.~Mebane, N.~Greiner, C.~Zhang, and S.~Willenbrock, ``{Constraints on
  Electroweak Effective Operators at One Loop},''
  \href{http://dx.doi.org/10.1103/PhysRevD.88.015028}{{\em Phys.Rev.}
  {\bfseries D88} no.~1, (2013) 015028},
\href{http://arxiv.org/abs/1306.3380}{{\ttfamily 1306.3380}}.

\bibitem{Jenkins:2013zja}
E.~E. Jenkins, A.~V. Manohar, and M.~Trott, ``{Renormalization Group Evolution
  of the Standard Model Dimension Six Operators I: Formalism and lambda
  Dependence},'' \href{http://dx.doi.org/10.1007/JHEP10(2013)087}{{\em JHEP}
  {\bfseries 10} (2013) 087},
\href{http://arxiv.org/abs/1308.2627}{{\ttfamily arXiv:1308.2627 [hep-ph]}}.

\bibitem{Jenkins:2013sda}
E.~E. Jenkins, A.~V. Manohar, and M.~Trott, ``{Naive Dimensional Analysis
  Counting of Gauge Theory Amplitudes and Anomalous Dimensions},''
  \href{http://dx.doi.org/10.1016/j.physletb.2013.09.020}{{\em Phys. Lett.}
  {\bfseries B726} (2013) 697--702},
\href{http://arxiv.org/abs/1309.0819}{{\ttfamily arXiv:1309.0819 [hep-ph]}}.

\bibitem{Jenkins:2013wua}
E.~E. Jenkins, A.~V. Manohar, and M.~Trott, ``{Renormalization Group Evolution
  of the Standard Model Dimension Six Operators II: Yukawa Dependence},''
  \href{http://dx.doi.org/10.1007/JHEP01(2014)035}{{\em JHEP} {\bfseries 01}
  (2014) 035},
\href{http://arxiv.org/abs/1310.4838}{{\ttfamily arXiv:1310.4838 [hep-ph]}}.

\bibitem{Alonso:2013hga}
R.~Alonso, E.~E. Jenkins, A.~V. Manohar, and M.~Trott, ``{Renormalization Group
  Evolution of the Standard Model Dimension Six Operators III: Gauge Coupling
  Dependence and Phenomenology},''
  \href{http://dx.doi.org/10.1007/JHEP04(2014)159}{{\em JHEP} {\bfseries 04}
  (2014) 159},
\href{http://arxiv.org/abs/1312.2014}{{\ttfamily arXiv:1312.2014 [hep-ph]}}.

\bibitem{Hartmann:2015oia}
C.~Hartmann and M.~Trott, ``{On one-loop corrections in the standard model
  effective field theory; the $\Gamma(h \rightarrow \gamma \, \gamma)$ case},''
  \href{http://dx.doi.org/10.1007/JHEP07(2015)151}{{\em JHEP} {\bfseries 07}
  (2015) 151},
\href{http://arxiv.org/abs/1505.02646}{{\ttfamily arXiv:1505.02646 [hep-ph]}}.

\bibitem{Ghezzi:2015vva}
M.~Ghezzi, R.~Gomez-Ambrosio, G.~Passarino, and S.~Uccirati, ``{NLO Higgs
  effective field theory and kappa-framework},''
  \href{http://dx.doi.org/10.1007/JHEP07(2015)175}{{\em JHEP} {\bfseries 07}
  (2015) 175},
\href{http://arxiv.org/abs/1505.03706}{{\ttfamily arXiv:1505.03706 [hep-ph]}}.

\bibitem{Zhang:2013xya}
C.~Zhang and F.~Maltoni, ``{Top-quark decay into Higgs boson and a light quark
  at next-to-leading order in QCD},''
  \href{http://dx.doi.org/10.1103/PhysRevD.88.054005}{{\em Phys. Rev.}
  {\bfseries D88} (2013) 054005},
\href{http://arxiv.org/abs/1305.7386}{{\ttfamily arXiv:1305.7386 [hep-ph]}}.

\bibitem{Englert:2014cva}
C.~Englert and M.~Spannowsky, ``{Effective Theories and Measurements at
  Colliders},'' \href{http://dx.doi.org/10.1016/j.physletb.2014.11.035}{{\em
  Phys. Lett.} {\bfseries B740} (2015) 8--15},
\href{http://arxiv.org/abs/1408.5147}{{\ttfamily arXiv:1408.5147 [hep-ph]}}.

\bibitem{Hartmann:2015aia}
C.~Hartmann and M.~Trott, ``{Higgs decay to two photons at one-loop in the
  SMEFT},''
\href{http://arxiv.org/abs/1507.03568}{{\ttfamily arXiv:1507.03568 [hep-ph]}}.

\bibitem{Cheung:2015aba}
C.~Cheung and C.-H. Shen, ``{Nonrenormalization Theorems without
  Supersymmetry},''
  \href{http://dx.doi.org/10.1103/PhysRevLett.115.071601}{{\em Phys. Rev.
  Lett.} {\bfseries 115} no.~7, (2015) 071601},
\href{http://arxiv.org/abs/1505.01844}{{\ttfamily arXiv:1505.01844 [hep-ph]}}.

\bibitem{Drozd:2015rsp}
A.~Drozd, J.~Ellis, J.~Quevillon, and T.~You, ``{The Universal One-Loop
  Effective Action},''
\href{http://arxiv.org/abs/1512.03003}{{\ttfamily arXiv:1512.03003 [hep-ph]}}.

\bibitem{Gauld:2015lmb}
R.~Gauld, B.~D. Pecjak, and D.~J. Scott, ``{One-loop corrections to $h\to b\bar
  b$ and $h\to \tau\bar \tau$ decays in the Standard Model Dimension-6 EFT:
  four-fermion operators and the large-$m_t$ limit},''
\href{http://arxiv.org/abs/1512.02508}{{\ttfamily arXiv:1512.02508 [hep-ph]}}.

\bibitem{Berthier:2015oma}
L.~Berthier and M.~Trott, ``{Towards consistent Electroweak Precision Data
  constraints in the SMEFT},''
  \href{http://dx.doi.org/10.1007/JHEP05(2015)024}{{\em JHEP} {\bfseries 05}
  (2015) 024},
\href{http://arxiv.org/abs/1502.02570}{{\ttfamily arXiv:1502.02570 [hep-ph]}}.

\bibitem{Christensen:2008py}
N.~D. Christensen and C.~Duhr, ``{FeynRules - Feynman rules made easy},''
  \href{http://dx.doi.org/10.1016/j.cpc.2009.02.018}{{\em Comput.Phys.Commun.}
  {\bfseries 180} (2009) 1614--1641},
\href{http://arxiv.org/abs/0806.4194}{{\ttfamily arXiv:0806.4194 [hep-ph]}}.

\bibitem{Alwall:2014hca}
J.~Alwall, R.~Frederix, S.~Frixione, V.~Hirschi, F.~Maltoni, O.~Mattelaer,
  H.~S. Shao, T.~Stelzer, P.~Torrielli, and M.~Zaro, ``{The automated
  computation of tree-level and next-to-leading order differential cross
  sections, and their matching to parton shower simulations},''
  \href{http://dx.doi.org/10.1007/JHEP07(2014)079}{{\em JHEP} {\bfseries 07}
  (2014) 079},
\href{http://arxiv.org/abs/1405.0301}{{\ttfamily arXiv:1405.0301 [hep-ph]}}.

\bibitem{Degrande:2011ua}
C.~Degrande, C.~Duhr, B.~Fuks, D.~Grellscheid, O.~Mattelaer, {\em et~al.},
  ``{UFO - The Universal FeynRules Output},''
  \href{http://dx.doi.org/10.1016/j.cpc.2012.01.022}{{\em Comput.Phys.Commun.}
  {\bfseries 183} (2012) 1201--1214},
\href{http://arxiv.org/abs/1108.2040}{{\ttfamily arXiv:1108.2040 [hep-ph]}}.

\bibitem{Campbell:2010ff}
J.~M. Campbell and R.~Ellis, ``{MCFM for the Tevatron and the LHC},''
  \href{http://dx.doi.org/10.1016/j.nuclphysbps.2010.08.011}{{\em
  Nucl.Phys.Proc.Suppl.} {\bfseries 205-206} (2010) 10--15},
\href{http://arxiv.org/abs/1007.3492}{{\ttfamily arXiv:1007.3492 [hep-ph]}}.

\bibitem{Frixione:2002ik}
S.~Frixione and B.~R. Webber, ``{Matching NLO QCD computations and parton
  shower simulations},''
  \href{http://dx.doi.org/10.1088/1126-6708/2002/06/029}{{\em JHEP} {\bfseries
  06} (2002) 029},
\href{http://arxiv.org/abs/hep-ph/0204244}{{\ttfamily arXiv:hep-ph/0204244
  [hep-ph]}}.

\bibitem{Frixione:2010wd}
S.~Frixione, F.~Stoeckli, P.~Torrielli, B.~R. Webber, and C.~D. White, ``{The
  MCaNLO 4.0 Event Generator},''
\href{http://arxiv.org/abs/1010.0819}{{\ttfamily arXiv:1010.0819 [hep-ph]}}.

\bibitem{Zhu:2001hw}
S.~Zhu, ``{Next-to-leading order QCD corrections to bg --> tW- at CERN large
  hadron collider},''
\href{http://arxiv.org/abs/hep-ph/0109269}{{\ttfamily arXiv:hep-ph/0109269
  [hep-ph]}}.

\bibitem{Campbell:2005bb}
J.~M. Campbell and F.~Tramontano, ``{Next-to-leading order corrections to Wt
  production and decay},''
  \href{http://dx.doi.org/10.1016/j.nuclphysb.2005.08.015}{{\em Nucl.Phys.}
  {\bfseries B726} (2005) 109--130},
\href{http://arxiv.org/abs/hep-ph/0506289}{{\ttfamily arXiv:hep-ph/0506289
  [hep-ph]}}.

\bibitem{Cao:2008af}
Q.-H. Cao, ``{Demonstration of One Cutoff Phase Space Slicing Method:
  Next-to-Leading Order QCD Corrections to the tW Associated Production in
  Hadron Collision},''
\href{http://arxiv.org/abs/0801.1539}{{\ttfamily arXiv:0801.1539 [hep-ph]}}.

\bibitem{Frixione:2008yi}
S.~Frixione, E.~Laenen, P.~Motylinski, B.~R. Webber, and C.~D. White,
  ``{Single-top hadroproduction in association with a W boson},''
  \href{http://dx.doi.org/10.1088/1126-6708/2008/07/029}{{\em JHEP} {\bfseries
  0807} (2008) 029},
\href{http://arxiv.org/abs/0805.3067}{{\ttfamily arXiv:0805.3067 [hep-ph]}}.

\bibitem{White:2009yt}
C.~D. White, S.~Frixione, E.~Laenen, and F.~Maltoni, ``{Isolating Wt production
  at the LHC},'' \href{http://dx.doi.org/10.1088/1126-6708/2009/11/074}{{\em
  JHEP} {\bfseries 0911} (2009) 074},
\href{http://arxiv.org/abs/0908.0631}{{\ttfamily arXiv:0908.0631 [hep-ph]}}.

\bibitem{Kauer:2001sp}
N.~Kauer and D.~Zeppenfeld, ``{Finite width effects in top quark production at
  hadron colliders},'' \href{http://dx.doi.org/10.1103/PhysRevD.65.014021}{{\em
  Phys.Rev.} {\bfseries D65} (2002) 014021},
\href{http://arxiv.org/abs/hep-ph/0107181}{{\ttfamily arXiv:hep-ph/0107181
  [hep-ph]}}.

\bibitem{Kersevan:2006fq}
B.~P. Kersevan and I.~Hinchliffe, ``{A Consistent prescription for the
  production involving massive quarks in hadron collisions},''
  \href{http://dx.doi.org/10.1088/1126-6708/2006/09/033}{{\em JHEP} {\bfseries
  09} (2006) 033},
\href{http://arxiv.org/abs/hep-ph/0603068}{{\ttfamily arXiv:hep-ph/0603068
  [hep-ph]}}.

\bibitem{Zhang:2010px}
C.~Zhang and S.~Willenbrock, ``{Effective Field Theory for Top Quark
  Physics},'' \href{http://dx.doi.org/10.1393/ncc/i2010-10678-9}{{\em Nuovo
  Cim.} {\bfseries C033} no.~4, (2010) 285--291},
\href{http://arxiv.org/abs/1008.3155}{{\ttfamily arXiv:1008.3155 [hep-ph]}}.

\bibitem{Czarnecki:2010gb}
A.~Czarnecki, J.~G. Korner, and J.~H. Piclum, ``{Helicity fractions of W bosons
  from top quark decays at NNLO in QCD},''
  \href{http://dx.doi.org/10.1103/PhysRevD.81.111503}{{\em Phys. Rev.}
  {\bfseries D81} (2010) 111503},
\href{http://arxiv.org/abs/1005.2625}{{\ttfamily arXiv:1005.2625 [hep-ph]}}.

\bibitem{Aaltonen:2011kc}
{\bfseries CDF} Collaboration, T.~Aaltonen {\em et~al.}, ``{Evidence for a Mass
  Dependent Forward-Backward Asymmetry in Top Quark Pair Production},''
  \href{http://dx.doi.org/10.1103/PhysRevD.83.112003}{{\em Phys. Rev.}
  {\bfseries D83} (2011) 112003},
\href{http://arxiv.org/abs/1101.0034}{{\ttfamily arXiv:1101.0034 [hep-ex]}}.

\bibitem{Czakon:2014xsa}
M.~Czakon, P.~Fiedler, and A.~Mitov, ``{Resolving the Tevatron Top Quark
  Forward-Backward Asymmetry Puzzle: Fully Differential
  Next-to-Next-to-Leading-Order Calculation},''
  \href{http://dx.doi.org/10.1103/PhysRevLett.115.052001}{{\em Phys. Rev.
  Lett.} {\bfseries 115} no.~5, (2015) 052001},
\href{http://arxiv.org/abs/1411.3007}{{\ttfamily arXiv:1411.3007 [hep-ph]}}.

\bibitem{Hollik:2011ps}
W.~Hollik and D.~Pagani, ``{The electroweak contribution to the top quark
  forward-backward asymmetry at the Tevatron},''
  \href{http://dx.doi.org/10.1103/PhysRevD.84.093003}{{\em Phys. Rev.}
  {\bfseries D84} (2011) 093003},
\href{http://arxiv.org/abs/1107.2606}{{\ttfamily arXiv:1107.2606 [hep-ph]}}.

\bibitem{Kuhn:2011ri}
J.~H. Kuhn and G.~Rodrigo, ``{Charge asymmetries of top quarks at hadron
  colliders revisited},'' \href{http://dx.doi.org/10.1007/JHEP01(2012)063}{{\em
  JHEP} {\bfseries 01} (2012) 063},
\href{http://arxiv.org/abs/1109.6830}{{\ttfamily arXiv:1109.6830 [hep-ph]}}.

\bibitem{Bernreuther:2012sx}
W.~Bernreuther and Z.-G. Si, ``{Top quark and leptonic charge asymmetries for
  the Tevatron and LHC},''
  \href{http://dx.doi.org/10.1103/PhysRevD.86.034026}{{\em Phys. Rev.}
  {\bfseries D86} (2012) 034026},
\href{http://arxiv.org/abs/1205.6580}{{\ttfamily arXiv:1205.6580 [hep-ph]}}.

\bibitem{Bauer:2010iq}
M.~Bauer, F.~Goertz, U.~Haisch, T.~Pfoh, and S.~Westhoff, ``{Top-Quark
  Forward-Backward Asymmetry in Randall-Sundrum Models Beyond the Leading
  Order},'' \href{http://dx.doi.org/10.1007/JHEP11(2010)039}{{\em JHEP}
  {\bfseries 11} (2010) 039},
\href{http://arxiv.org/abs/1008.0742}{{\ttfamily arXiv:1008.0742 [hep-ph]}}.

\bibitem{AguilarSaavedra:2011vw}
J.~A. Aguilar-Saavedra and M.~Perez-Victoria, ``{Probing the Tevatron t tbar
  asymmetry at LHC},'' \href{http://dx.doi.org/10.1007/JHEP05(2011)034}{{\em
  JHEP} {\bfseries 05} (2011) 034},
\href{http://arxiv.org/abs/1103.2765}{{\ttfamily arXiv:1103.2765 [hep-ph]}}.

\bibitem{Delaunay:2011gv}
C.~Delaunay, O.~Gedalia, Y.~Hochberg, G.~Perez, and Y.~Soreq, ``{Implications
  of the CDF $t \bar{t}$ Forward-Backward Asymmetry for Hard Top Physics},''
  \href{http://dx.doi.org/10.1007/JHEP08(2011)031}{{\em JHEP} {\bfseries 08}
  (2011) 031},
\href{http://arxiv.org/abs/1103.2297}{{\ttfamily arXiv:1103.2297 [hep-ph]}}.

\bibitem{AguilarSaavedra:2011cp}
J.~A. Aguilar-Saavedra, A.~Juste, and F.~Rubbo, ``{Boosting the $t \bar t$
  charge asymmetry},''
  \href{http://dx.doi.org/10.1016/j.physletb.2011.12.007}{{\em Phys. Lett.}
  {\bfseries B707} (2012) 92--98},
\href{http://arxiv.org/abs/1109.3710}{{\ttfamily arXiv:1109.3710 [hep-ph]}}.

\bibitem{Brehmer:2015rna}
J.~Brehmer, A.~Freitas, D.~Lopez-Val, and T.~Plehn, ``{Pushing Higgs Effective
  Theory to its Limits},''
\href{http://arxiv.org/abs/1510.03443}{{\ttfamily arXiv:1510.03443 [hep-ph]}}.

\bibitem{Isidori:2013cga}
G.~Isidori and M.~Trott, ``{Higgs form factors in Associated Production},''
  \href{http://dx.doi.org/10.1007/JHEP02(2014)082}{{\em JHEP} {\bfseries 02}
  (2014) 082},
\href{http://arxiv.org/abs/1307.4051}{{\ttfamily arXiv:1307.4051 [hep-ph]}}.

\bibitem{Boos:2006xe}
E.~Boos, V.~Bunichev, L.~Dudko, and M.~Perfilov, ``{Interference between
  $W^\prime$ and $W$ in single-top quark production processes},''
  \href{http://dx.doi.org/10.1016/j.physletb.2007.03.064}{{\em Phys. Lett.}
  {\bfseries B655} (2007) 245--250},
\href{http://arxiv.org/abs/hep-ph/0610080}{{\ttfamily arXiv:hep-ph/0610080
  [hep-ph]}}.

\bibitem{Aad:2011fq}
{\bfseries ATLAS} Collaboration, G.~Aad {\em et~al.}, ``{Search for New Physics
  in the Dijet Mass Distribution using 1 fb$^{-1}$ of $pp$ Collision Data at
  $\sqrt{s}=$7 TeV collected by the ATLAS Detector},''
  \href{http://dx.doi.org/10.1016/j.physletb.2012.01.035}{{\em Phys. Lett.}
  {\bfseries B708} (2012) 37--54},
\href{http://arxiv.org/abs/1108.6311}{{\ttfamily arXiv:1108.6311 [hep-ex]}}.

\bibitem{Aad:2014aqa}
{\bfseries ATLAS} Collaboration, G.~Aad {\em et~al.}, ``{Search for new
  phenomena in the dijet mass distribution using $p-p$ collision data at
  $\sqrt{s}=8$ TeV with the ATLAS detector},''
  \href{http://dx.doi.org/10.1103/PhysRevD.91.052007}{{\em Phys. Rev.}
  {\bfseries D91} no.~5, (2015) 052007},
\href{http://arxiv.org/abs/1407.1376}{{\ttfamily arXiv:1407.1376 [hep-ex]}}.

\bibitem{Khachatryan:2015sja}
{\bfseries CMS} Collaboration, V.~Khachatryan {\em et~al.}, ``{Search for
  resonances and quantum black holes using dijet mass spectra in proton-proton
  collisions at $\sqrt{s} =$ 8 TeV},''
  \href{http://dx.doi.org/10.1103/PhysRevD.91.052009}{{\em Phys. Rev.}
  {\bfseries D91} no.~5, (2015) 052009},
\href{http://arxiv.org/abs/1501.04198}{{\ttfamily arXiv:1501.04198 [hep-ex]}}.

\end{thebibliography}\endgroup
\end{document}